\documentclass[preprintnumbers,amsmath,11pt,amssymb,floatfix,superscriptaddress,nofootinbib]{article}

\def\mytitle#1{\setcounter{equation}{0}
\setcounter{footnote}{0}
\begin{flushleft}\Large\textbf{#1}\end{flushleft}
\vspace{0.25cm}}
\def\myname#1{\leftline{{\large #1}}\vspace{-0.13cm}}
\def\myplace#1#2{\small\begin{flushleft}\textit{#1}\\
\texttt{#2}\end{flushleft}}

\usepackage{graphicx}% Include figure files
\begin{document}

\mytitle{EMERGENT UNIVERSE WITH EXOTIC MATTER IN LOOP QUANTUM
COSMOLOGY, DGP BRANE WORLD AND KALUZA-KLEIN COSMOLOGY}

\vskip0.2cm \myname{ PRABIR
RUDRA~\footnote{prudra.math@gmail.com}}

\myplace{Department of Mathematics, Bengal Engineering and Science
University, Shibpur, Howrah-711 103, India.} {}

\begin{abstract}
In this work we have investigated the emergent scenario of the
universe described by Loop quantum cosmology model, DGP brane
model and Kaluza-Klein cosmology. Scalar field along with
barotropic fluid as normal matter is considered as the matter
content of the universe. In Loop quantum cosmology it is found
that the emergent scenario is realized with the imposition of some
conditions on the value of the density of normal matter in case of
normal and phantom scalar field. This is a surprising result
indeed considering the fact that scalar field is the dominating
matter component! In case of Tachyonic field, emergent scenario is
realized with some constraints on the value of $\rho_{1}$ for both
normal and phantom tachyon. In case of DGP brane-world realization
of an emergent scenario is possible almost unconditionally for
normal and phantom fields. Plots and table have been generated to
testify this fact. In case of tachyonic field emergent scenario is
realized with some constraints on $\dot{H}$. In Kaluza-Klein
cosmology emergent scenario is possible only for a closed universe
in case of normal and phantom scalar field. For a tachyonic field
realization of emergent universe is possible for all
models(closed, open and flat).
\end{abstract}

%%%%%%%%%%%%%%%%%%%%%%%%%%%%%%%%%%%%%%%%%%%%%%%%%%%%%%%%%%%%%%%%%%%%%%
\section{INTRODUCTION}
%%%%%%%%%%%%%%%%%%%%%%%%%%%%%%%%%%%%%%%%%%%%%%%%%%%%%%%%%%%%%%%%%%%%%%

\noindent

\vspace{3mm}

In the standard cosmological model, the existence of a big bang
singularity in the early universe is still an open problem. In
order to resolve this problem, Ellis et al. proposed, in the
context of general relativity, a scenario, called an emergent
universe \cite{Ellis1, Ellis2}. In this scenario, the space
curvature is positive and the Universe stays, past eternally, in
an Einstein static state and then evolves to a subsequent
inflationary phase. Inflation is basically future eternal, which
means that once inflation has started most of the volume of the
universe will remain in an inflating state. Probably this is the
possible mechanism responsible for the present cosmic acceleration
\cite{Perlmutter1, Spergel1}. So, the Universe originates from an
Einstein static state, rather than from a big bang singularity. It
is also worth noting that an Einstein static state as the initial
state of the Universe is also favored by the entropy
considerations \cite{Gibbons1, Gibbons2}.

\noindent

\vspace{3mm}

The inflationary universe emerges from a small static state that
has within it the seeds for the development of the microscopic
universe and it is called Emergent Universe scenario. The universe
has a finite initial size, with a finite amount of inflation
occurring over an infinite time in the past and with inflation
then coming to an end via reheating in the standard way.
References \cite{Mukherjee1, Debnath1} summarized the salient
features of emergent universe as follows:

\vspace{3mm}

1. The universe is almost static at the finite past ($t\rightarrow
-\infty$) and isotropic, homogeneous at large scales;\\

2. It is ever existing and there is no time-like singularity;\\

3. The universe is always large enough so that the classical
description of space-time is adequate;\\

4. The universe may contain exotic matter so that the energy
conditions may be violated;\\

5. The universe is accelerating as suggested by recent
measurements of distances of high redshift type Ia supernovae.\\

\noindent

\vspace{3mm}

Ellis et al \cite{Ellis2} provided a realization of a
singularity-free inflationary universe in the form of a simple
cosmological model dominated at early times by a single minimally
coupled scalar field with a physically based potential. Mukherjee
et al \cite{Mukherjee1} presented a general framework for an
emergent universe scenario and showed that emergent universe
scenarios are not isolated solutions and they may occur for
different combinations of radiation and matter. Campo et al
\cite{Campo1} studied the emergent universe model in the context
of a self-interacting Jordan-Brans-Dicke theory and showed that
the model presents a stable past eternal static solution which
eventually enters a phase where the stability of this solution is
broken leading to an inflationary period. Debnath \cite{Debnath1}
discussed the behaviour of different stages of the evolution of
the emergent universe considering that the universe is filled with
normal matter and a phantom field. Mukherji and Chakraborty
\cite{Mukherji1} developed Einstein-Gauss-Bonnet (EGB) theory in
the emergent universe scenario. They have considered the
Friedman-Lematre-Robertson-Walker cosmological model in Horava
gravity and the emergent scenario for all values of the spatial
curvature. Paul et al \cite{Paul1} predicted the range of the
permissible values for the parameters associated with the
constraints on exotic matter needed for an emergent universe. The
emergent universe in Horava gravity was studied by Mukherji and
Chakraborty \cite{Mukerji1}.

\noindent

\vspace{3mm}

In recent years, loop Quantum Gravity (LQG) is an outstanding
effort to describe the quantum effect of our universe
\cite{Rovelli1, Ashtekar1}. LQG is a theory trying to quantize the
gravity with a non-perturbative and background independent method.
The theory and principles of LQG when applied in the cosmological
framework creates a new theoretical framework of Loop Quantum
Cosmology(LQC) \cite{Ashtekar2, Bojowald1, Ashtekar3}. In this
theory, classical space-time continuum is replaced by a discrete
quantum geometry. The effect of LQG can be described by the
modification of Friedmann equation to add a term quadratic in
density. In LQC, the non-perturbative effects lead to correction
term $\frac{\rho_{T}^{2}}{\rho_{1}}$ to the standard Friedmann
equation. With the inclusion of this term, the universe bounces
quantum mechanically as the matter energy density reaches the
level of $\rho_{1}$(order of Plank density).

\noindent

\vspace{3mm}

A simple and effective model of brane-gravity is the
Dvali-Gabadadze-Porrati (DGP) braneworld model \cite{Dvali1} which
models our 4-dimensional world as a FRW brane embedded in a
5-dimensional Minkowski bulk. It explains the origin of dark
energy(DE) as the gravity on the brane leaking to the bulk at
large scale. On the 4-dimensional brane the action of gravity is
proportional to $M_p^2$ whereas in the bulk it is proportional to
the corresponding quantity in 5-dimensions. The model is then
characterized by a cross over length scale $
r_c=\frac{M_p^2}{2M_5^2} $ such that gravity is 4-dimensional
theory at scales $a<<r_c$ where matter behaves as pressureless
dust, but gravity leaks out into the bulk at scales $a>>r_c$ and
matter approaches the behaviour of a cosmological constant.
Moreover it has been shown that the standard Friedmann cosmology
can be firmly embedded in DGP brane.

\noindent

\vspace{3mm}

A fundamental theory on higher dimensional model was introduced by
Kaluza-Klein \cite{Kaluza1, Klein1}, where they considered an
extra dimension with FRW metric to unify Maxwell's theory of
electromagnetism and Einstein's gravity. After extensive research,
scientists realized that higher dimensional Cosmology may be more
useful to understand the interaction of particles. Moreover, as
our space-time is explicitly four dimensional in nature, the
'hidden' dimensions must be related to the dark matter and dark
energy. In modern physics Kaluza-Klein theory showed its great
importance in string theory \cite{Polchinski1}, in supergravity
\cite{Duff1} and in superstring theories \cite{Green1}. Li et al
in \cite{Li1} have considered the inflation in Kaluza-Klein
theory. In 1997, Overduin and Wesson \cite{Overduin1} represented
a review of Kaluza-Klein theory with higher dimensional unified
theories. String cloud and domain walls with quark matter in
N-dimensional Kaluza-Klein Cosmological model was presented by
Adhav et al \cite{Adhav1}. Recently some authors \cite{Qiang1,
Chen1, Ponce de Leon1, Chi1, Fukui1, Liu1, Coley1} studied
Kaluza-Klein cosmological models with different dark energies and
dark matters. In this paper we want to investigate the
possibilities for the realization of emergent universe for the
above three models.

\noindent

\vspace{3mm}

The general outline of the procedure followed in the paper is as
follows: At first we fix $H$ as a function of the energy density
and (possibly) pressure (this comes from the type of gravity we
choose); then given the Lagrangian of the field one knows the
energy density and pressure as functions of the field; finally, we
choose the cosmology to be of emergent type, which fixes H (or the
scale factor a) as function of time. The paper is organized as
follows: Section 2 deals with the emergent scenario in Loop
quantum cosmology. In section 3 we investigate the possibilities
for emergent scenario in DGP brane-world model. In section 4 we
discuss the emergent scenario for an universe described by
Kaluza-Klein cosmology. Finally the paper ends with some
concluding remarks in section 5.

%%%%%%%%%%%%%%%%%%%%%%%%%%%%%%%%%%%%%%%%%%%%%%%%%%%%%%%%%%%%%%%%%%%%%%
\section{EMERGENT SCENARIO IN LOOP QUANTUM COSMOLOGY}
%%%%%%%%%%%%%%%%%%%%%%%%%%%%%%%%%%%%%%%%%%%%%%%%%%%%%%%%%%%%%%%%%%%%%%

\noindent

Modified Friedmann equation for LQC is given by \cite{Wu1, Chen2,
Fu1}
\begin{equation}\label{emergent6.1}
H^{2}=\frac{\rho}{3}\left(1-\frac{\rho}{\rho_{1}}\right)
\end{equation}
Here $\rho_{1}=\sqrt{3}\pi^{2}\gamma^{3}G^{2}\hbar$ is the
critical loop quantum density and $\gamma$ is the dimensionless
Barbero-Immirzi parameter. The modified Raychaudhuri equation is
given by,
\begin{equation}\label{emergent6.2}
\dot{H}=-\frac{1}{2}\left(\rho+p\right)\left(1-2\frac{\rho}{\rho_{1}}\right)
\end{equation}

where $\rho=\rho_{\phi}+\rho_{m}$, is the total energy density
comprised of the energy density of normal matter, $\rho_{m}$ and
energy density of exotic matter, $\rho_{\phi}$, in the form of
scalar field.

\noindent

In this work we consider exotic matter in the form of phantom
field or tachyonic field and examine the possibility of an
emergent universe. Normal matter in the form of barotropic fluid
is considered to complement the exotic matter in a non-interacting
scenario. The pressure and energy density of normal matter is
connected by the relation,
\begin{equation}\label{emergent 6.2.1}
p_{m}=w\rho_{m},~~~~~~~-1\leq w \leq 1
\end{equation}
Now we consider that there is no interaction between normal matter
and phantom field (or tachyonic field), so the normal matter and
phantom field (or tachyonic field) are separately conserved. The
energy conservation equations for normal matter and phantom field
(or tachyonic field) are
\begin{equation}\label{emergent 6.2.2}
\dot{\rho_{m}}+3H\left(\rho_{m}+p_{m}\right)=0
\end{equation}
and

\begin{equation}\label{emergent 6.2.3}
\dot{\rho_{\phi}}+3H\left(\rho_{\phi}+p_{\phi}\right)=0
\end{equation}
From the eqn. (\ref{emergent 6.2.2}) we get the expression for
energy density of matter as

\begin{equation}\label{emergent 6.2.4}
\rho_{m}=\rho_{0}a^{-3\left(+w\right)}
\end{equation}
where $\rho_{0}$ is the integration constant.

\noindent

For emergent model of the universe, the scale factor may be chosen
as \cite{Mukherjee1, Debnath1}
\begin{equation}\label{emergent6.3}
a=a_{0}\left(\beta+e^{\alpha t}\right)^{n}
\end{equation}
where $a_{0}$, $\alpha$, $\beta$ and $n$ are positive constants.

\noindent

\vspace{2mm}

The constant parameters are restricted as follows \cite{Mukerji1}:
\vspace{3mm}

1. $a_{0}>0$ for the scale factor a to be positive, \vspace{3mm}

2. $\beta >0$, to avoid any singularity at finite time (big-rip)
\vspace{3mm}

3. $\alpha>0$ or $n>0$ for expanding model of the universe,
\vspace{3mm}

4. $\alpha<0$ and $n<0$ implies big bang singularity at
$t=-\infty$. \vspace{3mm}
~~~~~~~~~~~~~~~~~~~~~~~~~~~~~~~~~~~~~~~~~~~~~~~~~~~~~~~~~~~~~~~~~~~~~~~~~~~~~~~~~\\

\noindent

Therefore Hubble parameter and its derivatives are given by,
\begin{equation}\label{emergent6.4}
H=\frac{n\alpha e^{\alpha t}}{\left(\beta+e^{\alpha
t}\right)}~,~~\dot{H}=\frac{n\beta \alpha^{2}e^{\alpha
t}}{\left(\beta+e^{\alpha t}\right)^{2}}~, ~~\ddot{H}=\frac{n\beta
\alpha^{3}e^{\alpha t}\left(\beta-e^{\alpha
t}\right)}{\left(\beta+e^{\alpha t}\right)^{3}}
\end{equation}

\noindent

From the above expressions it is quite clear that $H$ and
$\dot{H}$ are both positive, but it is seen that $\ddot{H}$
changes sign at $t=\frac{1}{\alpha} \log\beta$. We see that $H$,
$\dot{H}$ and $\ddot{H}$ all tend to zero as
$t\rightarrow-\infty$. On the other hand as $t\rightarrow\infty$,
the solution gives asymptotically a de Sitter universe. From the
above choice of scale factor, the deceleration parameter $q$ can
be simplified to the form,
\begin{equation}\label{emergent6.5}
q=-1-\frac{\beta}{ne^{\alpha t}}
\end{equation}

%%%%%%%%%%%%%%%%%%%%%%%%%%%%%%%%%%%%%%%%%%%%%%%%%%%%%%%%%%%%%%%%%%
\subsection{NORMAL OR PHANTOM FIELD:}
%%%%%%%%%%%%%%%%%%%%%%%%%%%%%%%%%%%%%%%%%%%%%%%%%%%%%%%%%%%%%%%%%%%

\noindent

The kinetic energy term for phantom field has a negative sign. As
a result of this, the ratio between the pressure and energy
density is always less than $-1$. The explicit form of energy
density and pressure are as follows, \cite{Chang1}
\begin{equation}\label{emergent6.6}
\rho_{\phi}=\frac{\delta}{2}\dot{\phi}^{2}+V\left(\phi\right)
\end{equation}

\begin{equation}\label{emergent6.7}
p_{\phi}=\frac{\delta}{2}\dot{\phi}^{2}-V\left(\phi\right)
\end{equation}
where $\delta=+1$ corresponds to the normal scalar field  and
$\delta=-1$ corresponds to the phantom scalar field. Now using
equations (\ref{emergent6.2})~,~ (\ref{emergent6.6}) and
(\ref{emergent6.7}) we have,
\begin{equation}\label{emergent6.8}
\dot{H}=-\frac{1}{2\rho_{1}}\left[\delta\dot{\phi}^{2}+\left(1+w\right)\rho_{m}\right]\left[\rho_{1}-\delta\dot{\phi}^{2}-2V(\phi)-2\rho_{m}\right]
\end{equation}

\noindent

This is a quadratic in $\dot{\phi}^{2}$ and we get,
\begin{equation}\label{emergent6.9}
\dot{\phi}^{2}=\frac{-B+\sqrt{B^{2}-4AC}}{2A}
\end{equation}
where $A=\frac{\delta^{2}}{2\rho_{1}}$,
~$B=\frac{\delta}{2\rho_{1}}\left[\left(w+3\right)\rho_{m}-\rho_{1}+2V(\phi)\right]$
and
$C=-\left[\frac{\rho_{m}}{2\rho_{1}}\{\left(1+w\right)\left(\rho_{1}-2V(\phi)\right)-2\}-\dot{H}\right]$.

%%%%%%%%%%%%%%%%%%%%%%%%%%%%%%%%%%%%%%%%%%%%%%%%%%%%%%%%%%%%%%%%
\subsubsection{Case I : For Normal scalar field}
%%%%%%%%%%%%%%%%%%%%%%%%%%%%%%%%%%%%%%%%%%%%%%%%%%%%%%%%%%%%%%%%

\noindent

\vspace{3mm}

For normal scalar field, $\delta=+1$. Therefore
$B=\frac{1}{2\rho_{1}}\left[\left(w+3\right)\rho_{m}-\rho_{1}+2V(\phi)\right]$.\\

We have two cases:

\vspace{3mm}
~~~~~~~~~~~~~~~~~~~~~~~~~~~~~~~~~~~~~~~~~~~~~~~~~~~~~~~~~~~~~~~~~~~~~~~~~~~~~~~~~~~~~~~~~~\\
~~~~~~~~~~~~~~~~~~~~~~~~~~~~~~~~~~~~~~~~~~~~~~~~~~~~~~~~~~~~~~~~~~~~~~~~~~~~~~~~~~~~~~~~~~~\\
1) for $B<0$ : We have
$V\left(\phi\right)<\frac{1}{2}\left\{\rho_{1}-\left(w+3\right)\rho_{m}\right\}$.
Now we should have $B^{2}-4AC>0$. Since $A>0$, then $C<0$.
Therefore we have
$\dot{H}<\frac{\rho_{m}}{2\rho_{1}}\left\{\left(1+w\right)\left(\rho_{1}-2V(\phi)\right)-2\right\}$.
Now in order to realize a universe undergoing accelerated
expansion we should have $\dot{H}>0$. This is possible if
$\rho_{1}-2V(\phi)-2>0$, i.e.,
$V(\phi)<\frac{1}{2}\left(\rho_{1}-2\right)$. Now comparing this
with the accepted range of $V(\phi)$, we get
$w<\left(\frac{2}{\rho_{m}}-3\right)<1$, which implies
$\rho_{m}>\frac{1}{2}$. This is a very interesting result.
\vspace{3mm}
~~~~~~~~~~~~~~~~~~~~~~~~~~~~~~~~~~~~~~~~~~~~~~~~~~~~~~~~~~~~~~~~~~~~~~~~~~~~~~~~~~~~~~~~~~~~\\
~~~~~~~~~~~~~~~~~~~~~~~~~~~~~~~~~~~~~~~~~~~~~~~~~~~~~~~~~~~~~~~~~~~~~~~~~~~~~~~~~~~~~~~~~~~~\\
2) for $B>0$ : We have
$V\left(\phi\right)>\frac{1}{2}\left\{\rho_{1}-\left(w+3\right)\rho_{m}\right\}$.
Now in order to make the R.H.S. of equation (\ref{emergent6.9})
positive definite, we should have $C<0$, which again leads us to
the result $V(\phi)<\frac{1}{2}\left(\rho_{1}-2\right)$ for the
accelerating universe just like the previous case. Now if we
compare this with the accepted range for $V(\phi)$, we get a range
for the values of the potential,
$\frac{1}{2}\{\rho_{1}-\left(w+3\right)\rho_{m}\}<V(\phi)<\frac{1}{2}\left(\rho_{1}-2\right)$.
From this we further get $w>\left(\frac{2}{\rho_{m}}-3\right)>-1$,
which implies $\rho_{m}<1$. This is also a very important result.

\vspace{3mm}

%%%%%%%%%%%%%%%%%%%%%%%%%%%%%%%%%%%%%%%%%%%%%%%%%%%%%%%%%%%%%%%%%%%%
\subsubsection{Case II : For Phantom field}
%%%%%%%%%%%%%%%%%%%%%%%%%%%%%%%%%%%%%%%%%%%%%%%%%%%%%%%%%%%%%%%%%%%%

\noindent

For phantom field, $\delta=-1$. Therefore
$B=-\frac{1}{2\rho_{1}}\left[\left(w+3\right)\rho_{m}-\rho_{1}+2V(\phi)\right]$.\\\\

Like before we have two cases:

\vspace{3mm}
~~~~~~~~~~~~~~~~~~~~~~~~~~~~~~~~~~~~~~~~~~~~~~~~~~~~~~~~~~~~~~~~~~~~~~~~~~~~~~~~~~~~~~~~~~\\
~~~~~~~~~~~~~~~~~~~~~~~~~~~~~~~~~~~~~~~~~~~~~~~~~~~~~~~~~~~~~~~~~~~~~~~~~~~~~~~~~~~~~~~~~~~\\
1) for $B<0$ : We have
$V\left(\phi\right)>\frac{1}{2}\{\rho_{1}-\left(w+3\right)\rho_{m}\}$.
Now we should have $B^{2}-4AC>0$. Hence this is just like sub-case
2 of normal scalar field. We get a range for the potential
$V(\phi)$,
$\frac{1}{2}\{\rho_{1}-\left(w+3\right)\rho_{m}\}<V(\phi)<\frac{1}{2}\left(\rho_{1}-2\right)$.
From this we further get $w>\left(\frac{2}{\rho_{m}}-3\right)>-1$,
which implies $\rho_{m}<1$ for the accelerating universe.

\vspace{3mm}

~~~~~~~~~~~~~~~~~~~~~~~~~~~~~~~~~~~~~~~~~~~~~~~~~~~~~~~~~~~~~~~~~~~~~~~~~~~~~~~~~~~~~~~~~~~~\\
~~~~~~~~~~~~~~~~~~~~~~~~~~~~~~~~~~~~~~~~~~~~~~~~~~~~~~~~~~~~~~~~~~~~~~~~~~~~~~~~~~~~~~~~~~~~\\
2) for $B>0$ : We have
$V\left(\phi\right)<\frac{1}{2}\{\rho_{1}-\left(w+3\right)\rho_{m}\}$.
Now in order to make the R.H.S. of equation (\ref{emergent6.9})
positive definite, we should have $C<0$, i.e., $\dot{H}>0$. This
again leads to sub-case 1 of the normal scalar field. Hence we
should have $\rho_{1}-2V(\phi)-2>0$, i.e.,
$V(\phi)<\frac{1}{2}\left(\rho_{1}-2\right)$ for the accelerating
universe. Now comparing this with the accepted range of $V(\phi)$,
we get $w<\left(\frac{2}{\rho_{m}}-3\right)<1$, which implies
$\rho_{m}>\frac{1}{2}$.

\vspace{3mm}

%%%%%%%%%%%%%%%%%%%%%%%%%%%%%%%%%%%%%%%%%%%%%%%%%%%%%%%%%%%%%%
\subsection{TACHYONIC FIELD}
%%%%%%%%%%%%%%%%%%%%%%%%%%%%%%%%%%%%%%%%%%%%%%%%%%%%%%%%%%%%%%

\noindent

\vspace{3mm}

The Lagrangian for the tachyonic field $\psi$ having potential
$U\left(\psi\right)$ is given by \cite{Hao1, Hao2, Nojiri1,
Gumjudpai1},
\begin{equation}\label{emergent6.11}
L=-U\left(\psi\right)\sqrt{1-\epsilon\dot{\psi}^{2}}
\end{equation}
where $\epsilon=+1$ represents the normal tachyon and
$\epsilon=-1$ represents phantom tachyon. The expressions for
energy density and pressure are,
\begin{equation}\label{emergent6.12}
\rho_{\psi}=\frac{U\left(\psi\right)}{\sqrt{1-\epsilon\dot{\psi}^{2}}}
\end{equation}
and,
\begin{equation}\label{emergent6.13}
p_{\psi}=-U\left(\psi\right)\sqrt{1-\epsilon\dot{\psi}^{2}}
\end{equation}

\noindent

Due to the above complicated forms of $\rho$ and $p$, it is not
possible to find an expression for $\dot{\psi}^{2}$. However using
the field equation for LQC, i.e., equation (\ref{emergent6.2}), we
get,
\begin{equation}\label{emergent6.14}
\dot{\psi}^{2}=\frac{2\rho_{m}\left(1+w\right)\left(\rho_{\psi}+\rho_{m}\right)-\rho_{1}\rho_{m}\left(1+w\right)-2\dot{H}\rho_{1}}{\rho_{1}\rho_{\psi}\epsilon-2\epsilon\rho_{\psi}\left(\rho_{\psi}+\rho_{m}\right)}
\end{equation}
We know that $\dot{\psi}^{2}$ has to be positive. It means that
both numerator and denominator are of the same sign.

\noindent

\vspace{5mm}

{\bf Case I: When both numerator and denominator are positive:} \\

For phantom tachyon, $\epsilon=-1$. We get $\rho_{1}<2\rho$ and
$\rho_{1}<2\rho\left[\frac{\rho_{m}\left(1+w\right)}{\rho_{m}\left(1+w\right)+2\dot{H}}\right]$.
We know that for an accelerating universe $\dot{H}>0$. So keeping
this in mind we combine the above two inequalities into a single
one
$\rho_{1}<2\rho\left[\frac{\rho_{m}\left(1+w\right)}{\rho_{m}\left(1+w\right)+2\dot{H}}\right]$,
i.e., $\rho_{1}\in
\left(0,2\rho\left[\frac{\rho_{m}\left(1+w\right)}{\rho_{m}\left(1+w\right)+2\dot{H}}\right]\right)$
since the second one implies the first.

\vspace{5mm}

\noindent

For normal phantom, $\epsilon=1$. Here we get $\rho_{1}>2\rho$ and
$\rho_{1}<2\rho\left[\frac{\rho_{m}\left(1+w\right)}{\rho_{m}\left(1+w\right)+2\dot{H}}\right]$.
It is quite clear that for an accelerating universe the acceptable
range of values for $\rho_{1}$ is $\rho_{1}\in
\left(0,2\rho\left[\frac{\rho_{m}\left(1+w\right)}{\rho_{m}\left(1+w\right)+2\dot{H}}\right]\right)
\bigcup\left(2\rho,\infty\right)$

\vspace{7mm}

{\bf Case II: When both numerator and denominator are negative:} \\

For phantom tachyon, we get $\rho_{1}>2\rho$ and
$\rho_{1}>2\rho\left[\frac{\rho_{m}\left(1+w\right)}{\rho_{m}\left(1+w\right)+2\dot{H}}\right]$.
Combining the two inequalities for the accelerating universe we
get the acceptable range of $\rho_{1}$ as $\rho_{1}>2\rho$, i.e.,
$\rho_{1}\in \left(2\rho, \infty\right)$

\vspace{5mm}

\noindent

For normal tachyon we get $\rho_{1}<2\rho$ and
$\rho_{1}>2\rho\left[\frac{\rho_{m}\left(1+w\right)}{\rho_{m}\left(1+w\right)+2\dot{H}}\right]$.
So combining the above two we get the acceptable range for
$\rho_{1}$ as
$2\rho\left[\frac{\rho_{m}\left(1+w\right)}{\rho_{m}\left(1+w\right)+2\dot{H}}\right]<\rho_{1}<2\rho$,
i.e., $\rho_{1}\in
\left(2\rho\left[\frac{\rho_{m}\left(1+w\right)}{\rho_{m}\left(1+w\right)+2\dot{H}}\right],
2\rho\right)$

\vspace{5mm}

%%%%%%%%%%%%%%%%%%%%%%%%%%%%%%%%%%%%%%%%%%%%%%%%%%%%%%%%%%%%%%%%%%%%%
\section{EMERGENT SCENARIO IN DGP BRANE-WORLD}
%%%%%%%%%%%%%%%%%%%%%%%%%%%%%%%%%%%%%%%%%%%%%%%%%%%%%%%%%%%%%%%%%%%%%

\noindent

\vspace{3mm}

While flat, homogeneous and isotropic brane is being considered,
the Friedmann equation in DGP brane model \cite{Dvali1, Deffayet1,
Deffayet2} is modified to the equation
\begin{equation}\label{emergent6.15}
H^2=\left(\sqrt{\frac{\rho}{3}+\frac{1}{4r_{c}^{2}}}+\epsilon'
\frac{1}{2r_c}\right)^2,
\end{equation}
where $H=\frac{\dot a}{a}$ is the Hubble parameter, $\rho$ is the
total cosmic fluid energy density and $r_c=\frac{M_p^2}{2M_5^2}$
is the cross-over scale which determines the transition from 4D to
5D behaviour and $\epsilon'=\pm 1 $ (choosing $M_{p}^{2}=8\pi
G=1$). For $\epsilon'=+1$, we have standard DGP$(+)$ model which
is self accelerating model without any form of DE, and effective
$w$ is always non-phantom. However for $\epsilon'=-1$, we have
DGP$(-)$ model which does not self accelerate but requires DE on
the brane. It experiences 5D gravitational modifications to its
dynamics which effectively screen DE. The modified Raychaudhuri's
equation is given by
\begin{equation}\label{emergent6.16}
\left(2H-\frac{\epsilon'}{r_{c}}\right)\dot{H}=-H\left(\rho+p\right)
\end{equation}

%%%%%%%%%%%%%%%%%%%%%%%%%%%%%%%%%%%%%%%%%%%%%%%%%%%%%%%%%%%%%%%%%%
\subsection{NORMAL OR PHANTOM FIELD:}
%%%%%%%%%%%%%%%%%%%%%%%%%%%%%%%%%%%%%%%%%%%%%%%%%%%%%%%%%%%%%%%%%%%

\noindent

\vspace{3mm}

The relevant expressions for $\rho$ and $p$ are given by equations
(\ref{emergent6.6}) and (\ref{emergent6.7}) respectively. Now
using equations (\ref{emergent6.15}), (\ref{emergent6.16}) and
using the values of $\rho$ and $p$ from equations
(\ref{emergent6.6}) and (\ref{emergent6.7}) we get,

$$\frac{1}{r}\left[\left\{\rho_{m}\left(w+1\right)+\delta\dot{\phi}^{2}\right\}^{2}\left\{3\left(1+\epsilon'^{2}\right)+2r_{c}^{2}\left(2V(\phi)+2\rho_{m}+\delta\dot{\phi}^{2}\right)+2r_{c}\sqrt{3}\epsilon'\sqrt{\frac{3}{r_{c}^{2}}+4V(\phi)+4\rho_{m}+2\delta\dot{\phi}^{2}}\right\}\right.$$
\begin{equation}\label{emergent6.17}
\left.-4\left\{3+2r_{c}^{2}\left(2V(\phi)+2\rho_{m}+\delta\dot{\phi}^{2}\right)\right\}\dot{H}^{2}\right]=0
\end{equation}

\noindent

\vspace{3mm}

This is a $8$ degree equation in $\dot{\phi}$. It is not possible
to find an explicit expression for $\dot{\phi}$ in terms of the
given parameters. So we intend to find a numerical solution of the
above equation by considering different sets of values of the
parameters involved. The following plots and the table show the
possible outcomes of the procedure undertaken and the positive
solutions of $\dot{\phi}$ obtained for different sets of values of
the parameters.

\noindent

\vspace{3mm}

\begin{figure}\label{figure}
\includegraphics[height=2in, width=2in]{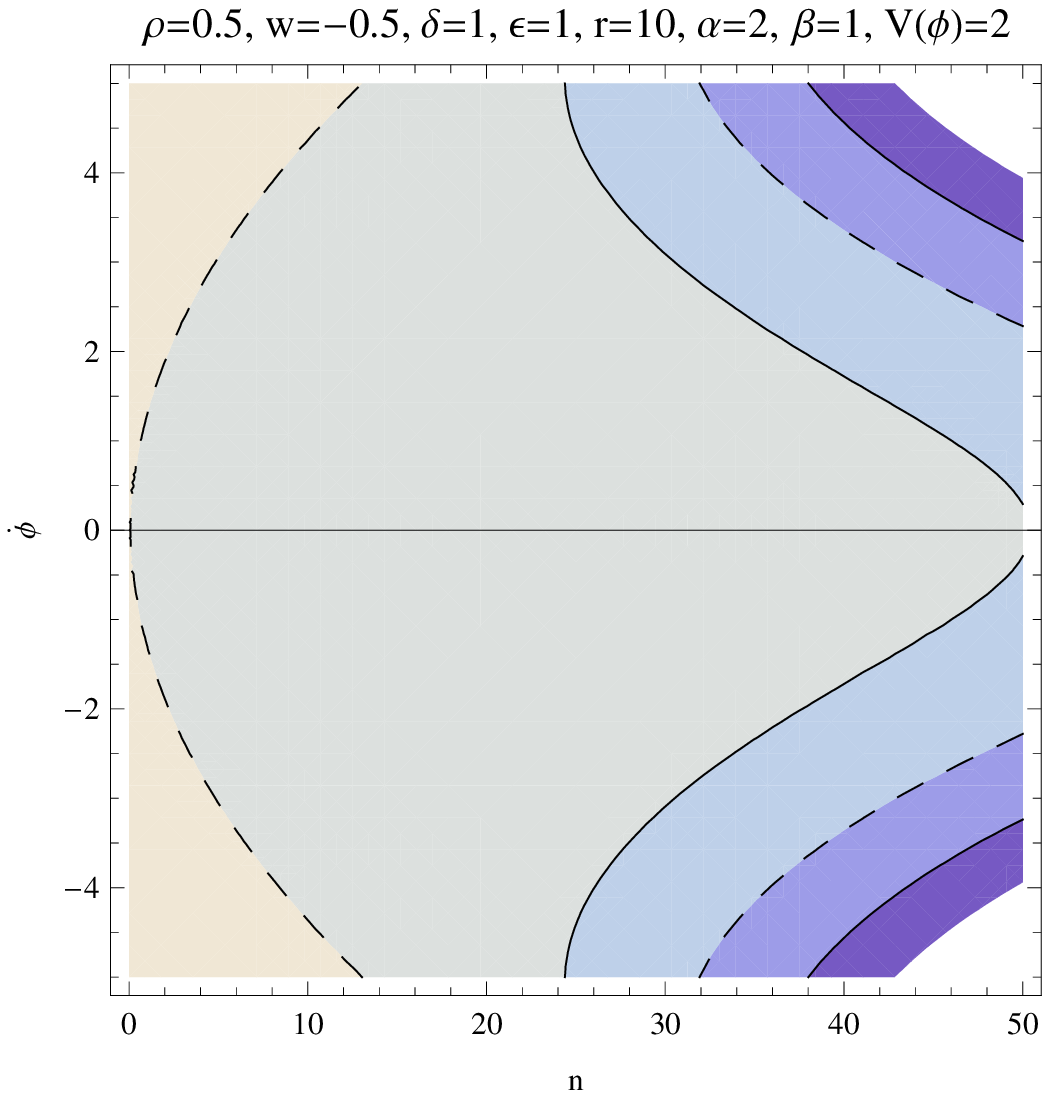}~~\includegraphics[height=2in, width=2in]{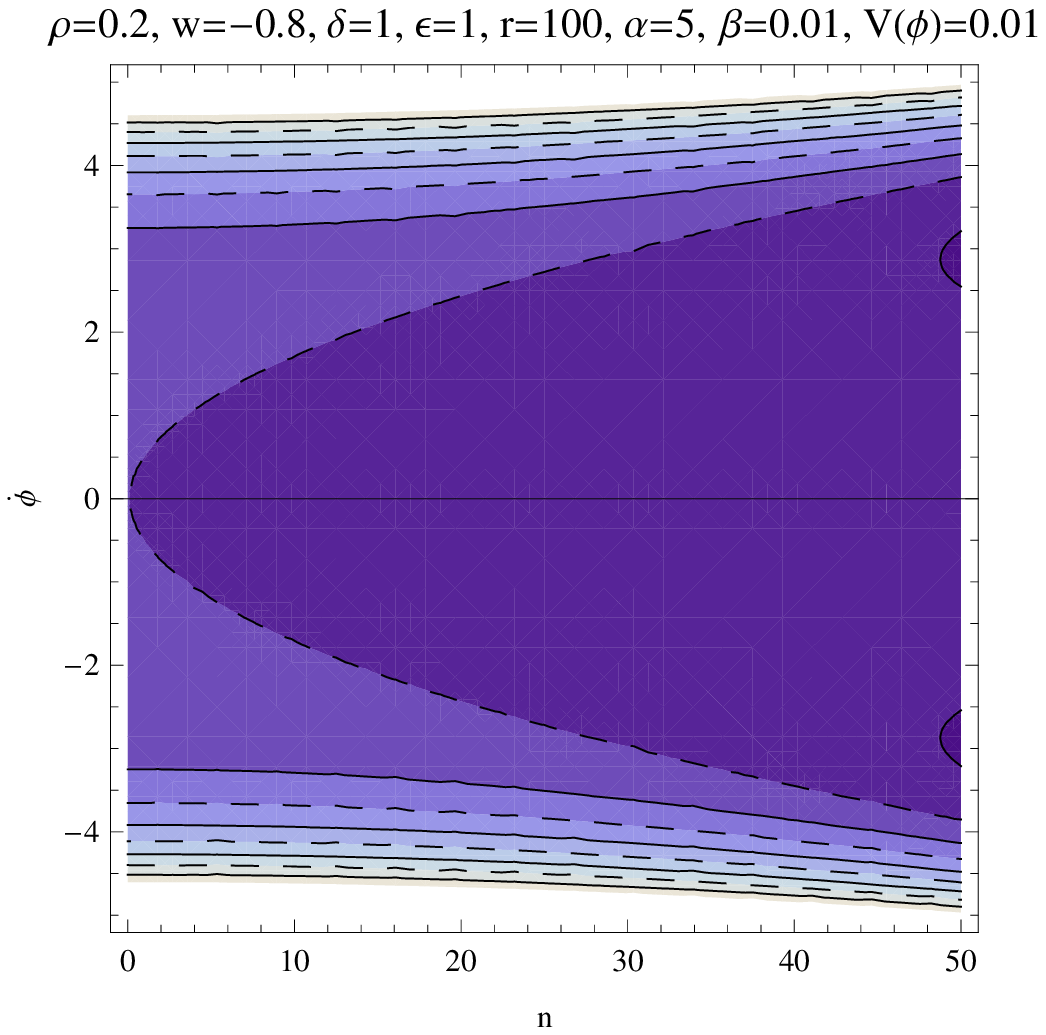}~~\includegraphics[height=2in, width=2in]{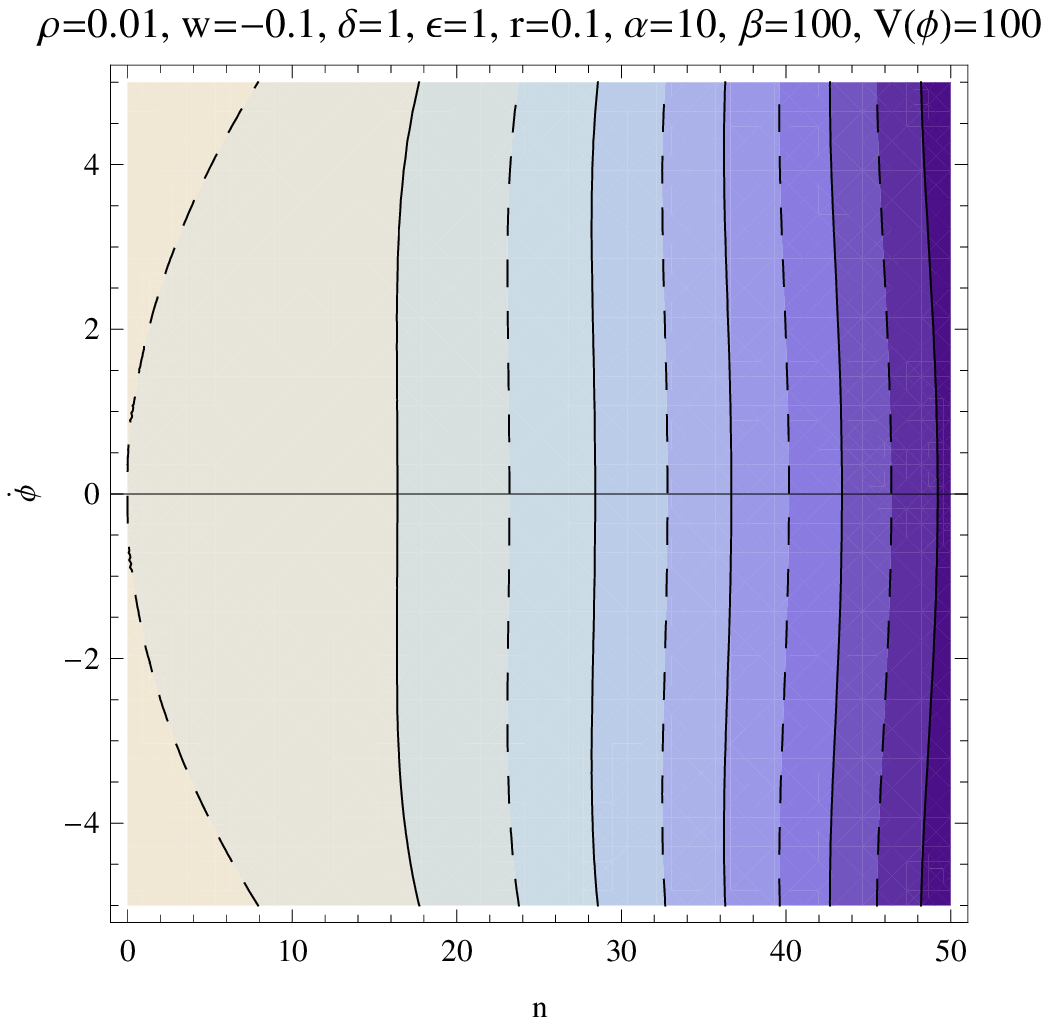}\\
\vspace{3mm}
~~~~~~~~~~~~~~Fig. 1~~~~~~~~~~~~~~~~~~~~~~~~~~~~~~~~~~Fig. 2~~~~~~~~~~~~~~~~~~~~~~~~~~~~~~~~~~Fig. 3~~~~~~~~~~~\\
\includegraphics[height=2in, width=2in]{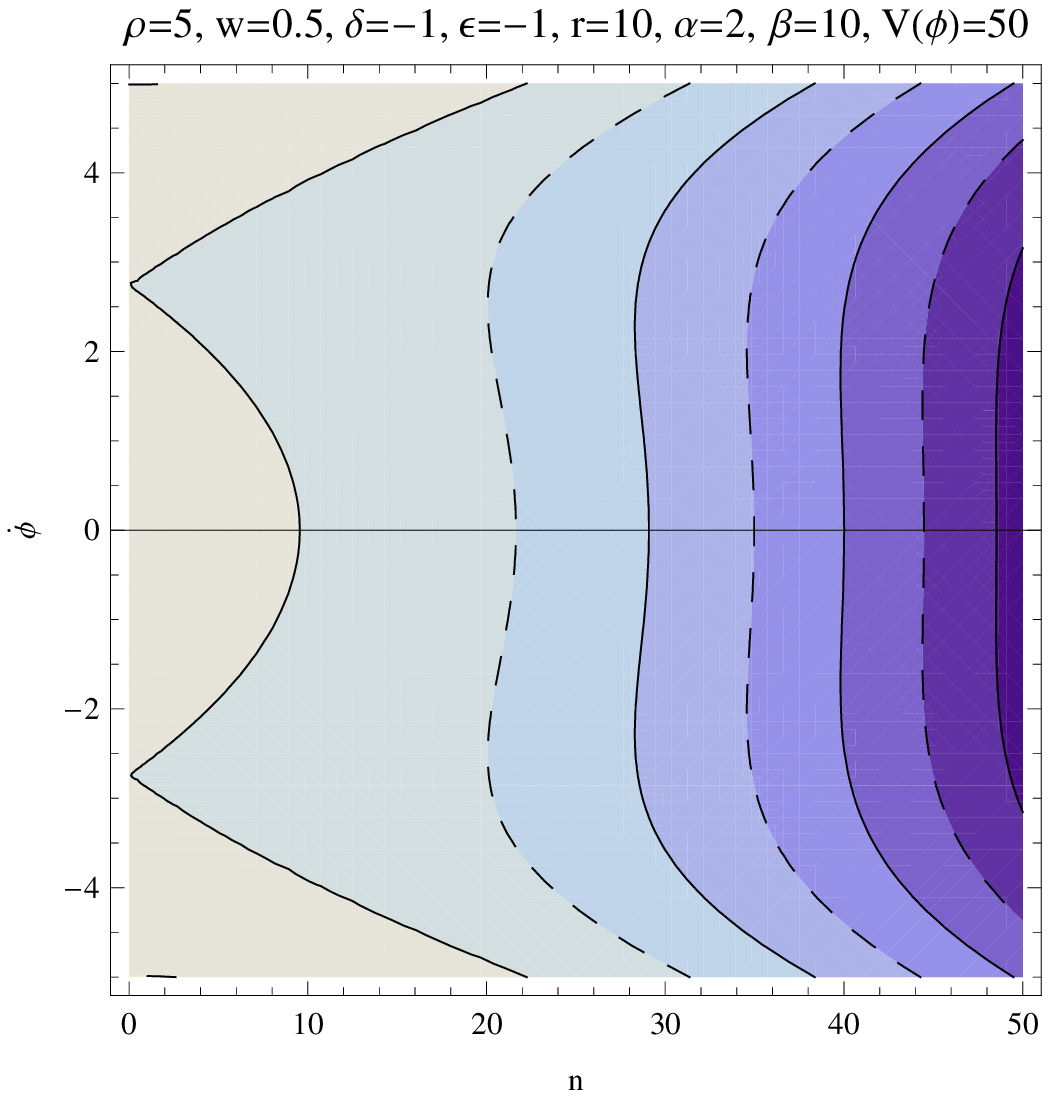}~~\includegraphics[height=2in, width=2in]{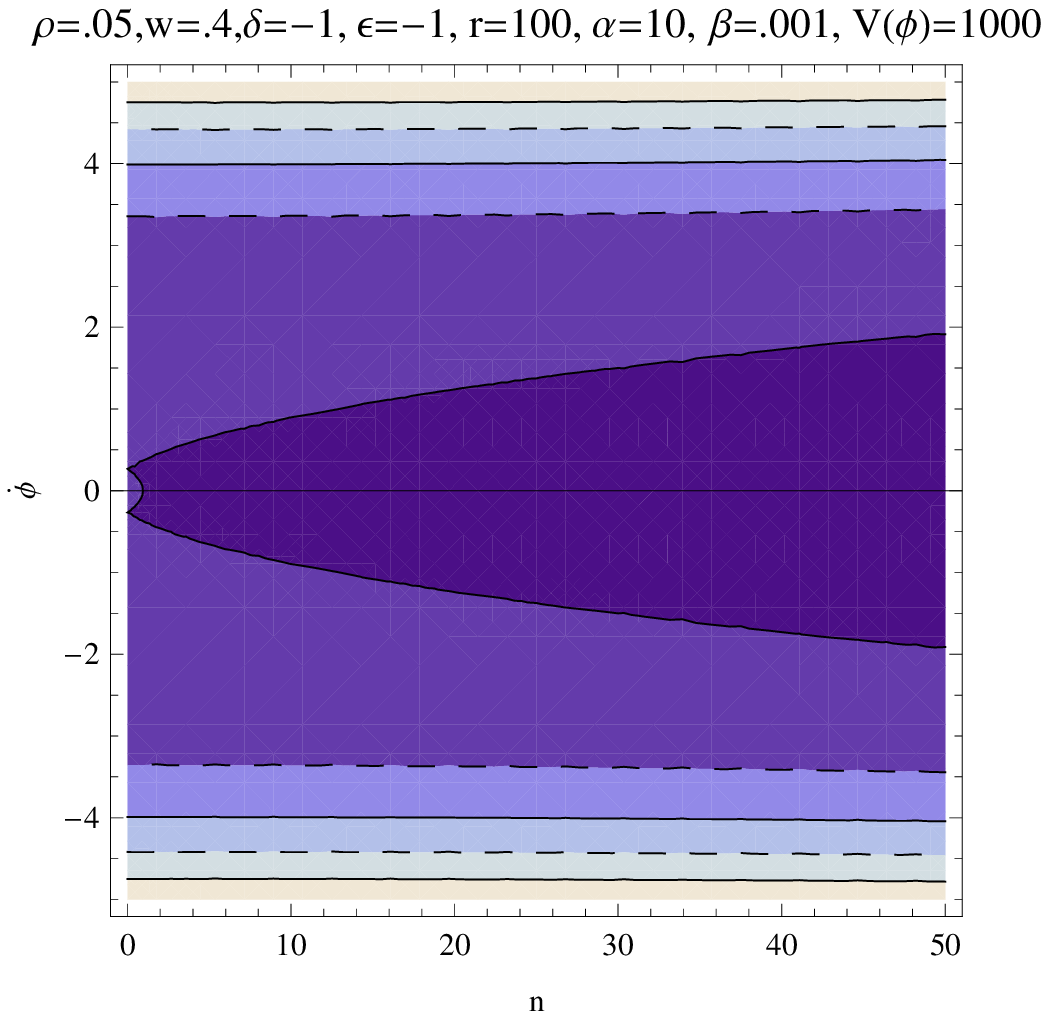}~~\includegraphics[height=2in, width=2in]{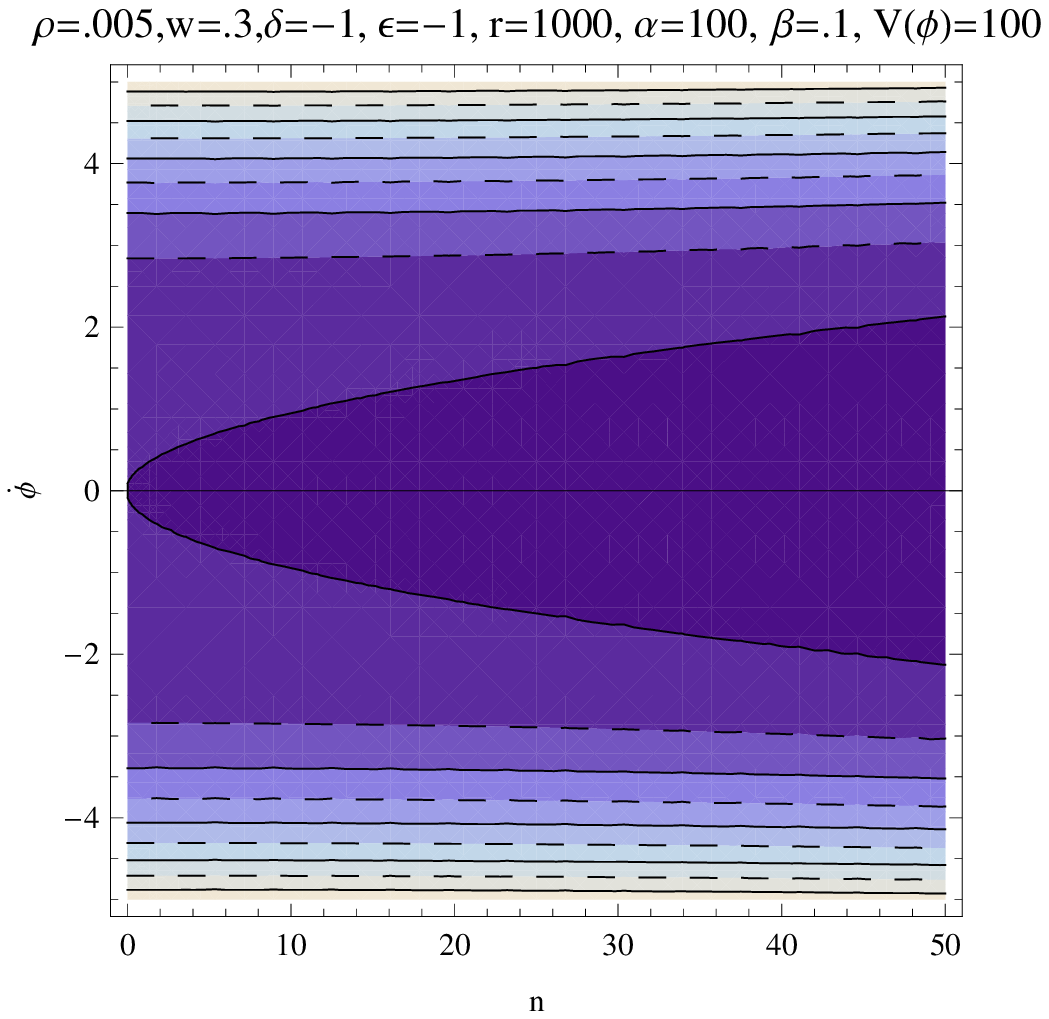}\\
\vspace{3mm}
~~~~~~~~~~~~~~Fig. 4~~~~~~~~~~~~~~~~~~~~~~~~~~~~~~~~~~Fig. 5~~~~~~~~~~~~~~~~~~~~~~~~~~~~~~~~~~Fig. 6~~~~~~~~~~~\\
\includegraphics[height=2in, width=2in]{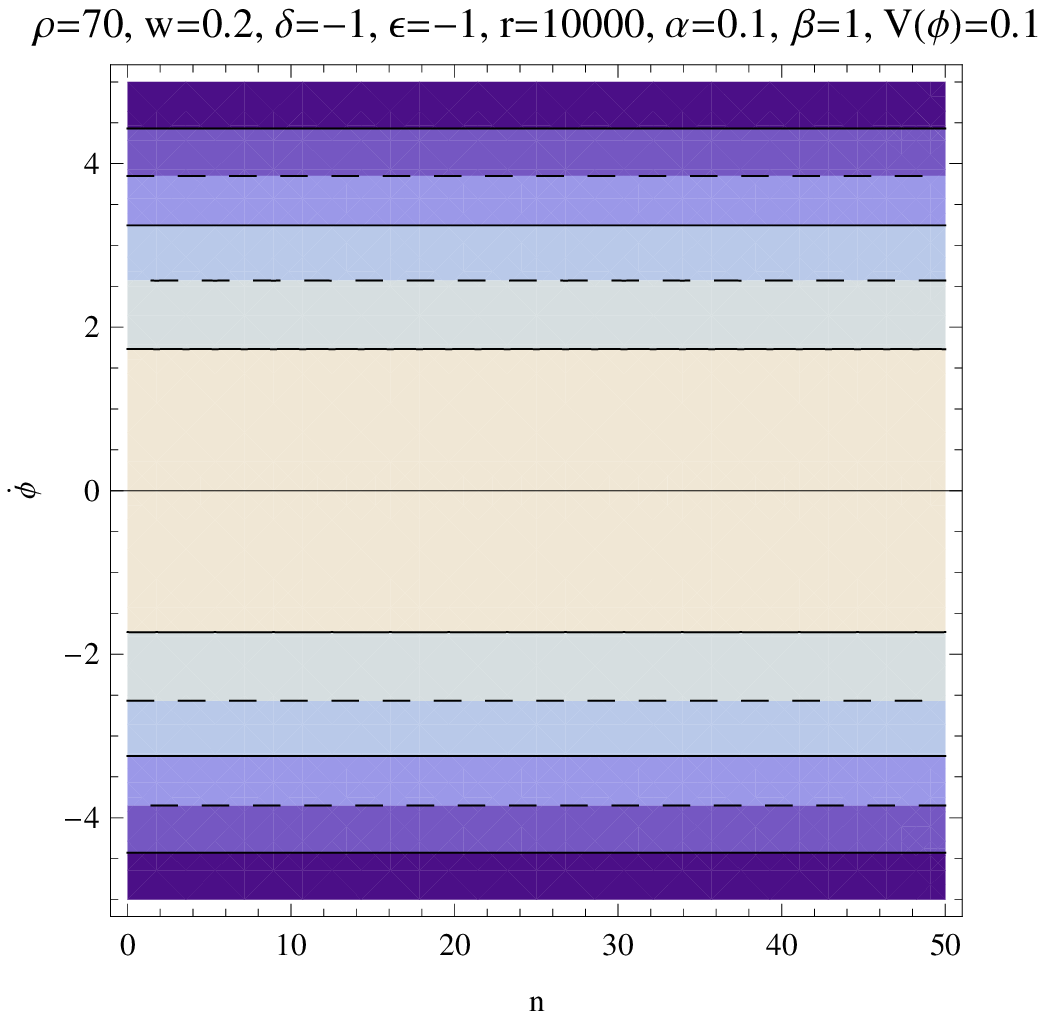}~~\includegraphics[height=2in, width=2in]{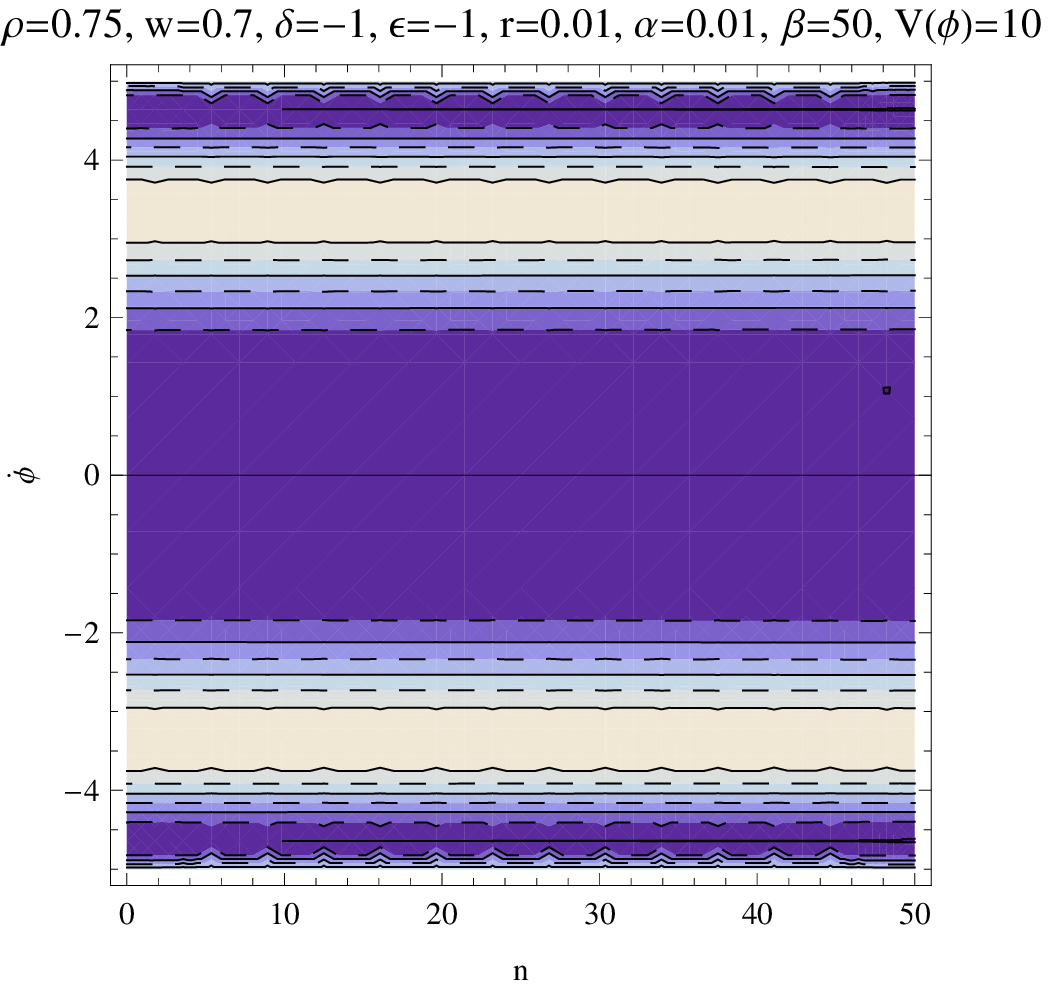}~~\\
\vspace{3mm}
~~~~~~~~~~~~~~Fig. 7~~~~~~~~~~~~~~~~~~~~~~~~~~~~~~~~~~Fig. 8~~~\\

The above figures represent the plot of $\dot{\phi}$ against $n$
(in equation (20)) with the variation of different parameters. The
values of the parameters chosen are provided with the plots.
\end{figure}

\begin{figure}
\includegraphics[height=2in, width=2in]{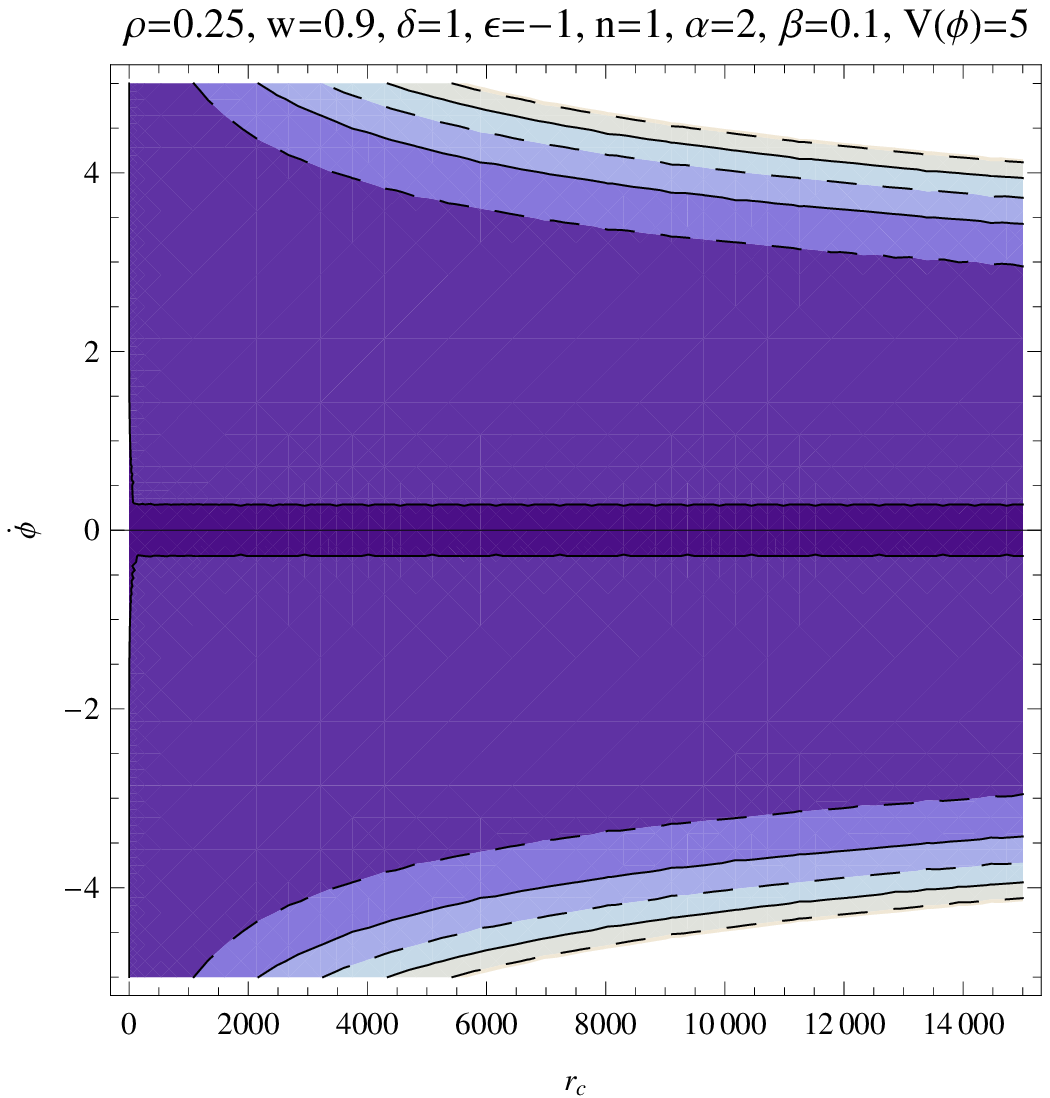}\includegraphics[height=2in, width=2in]{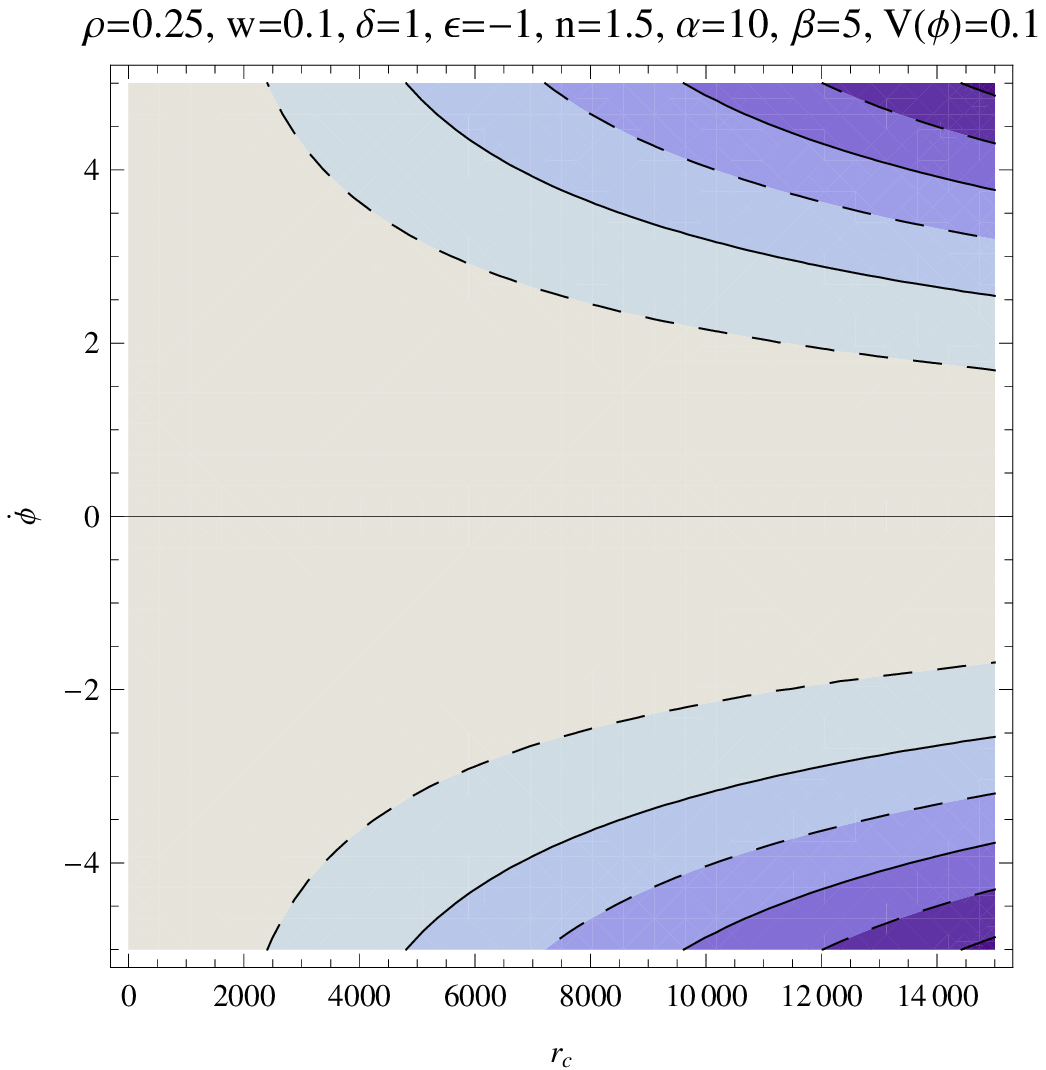}~~\includegraphics[height=2in, width=2in]{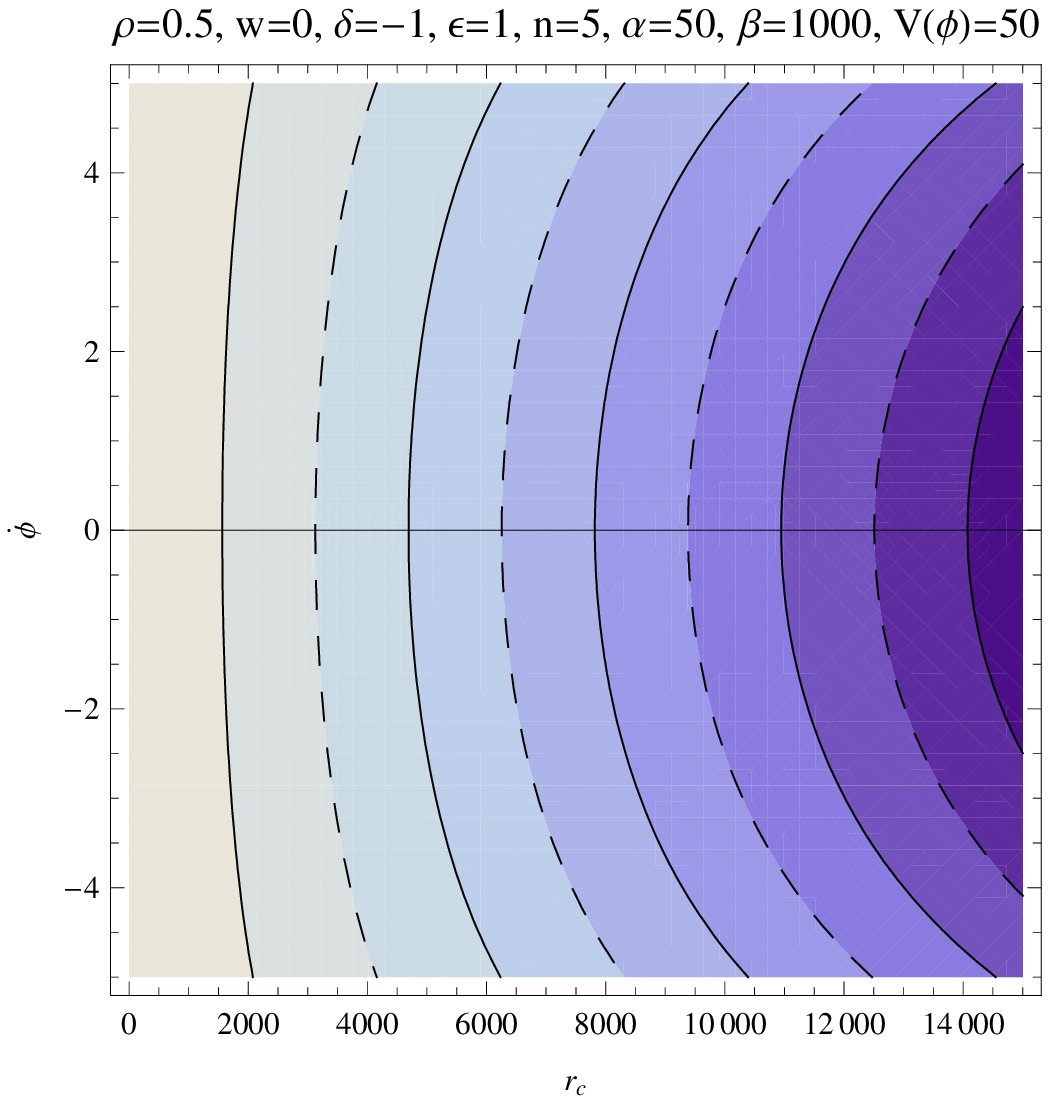}~~\\
\vspace{3mm}
~~~~~~~~~~~~~~Fig. 9~~~~~~~~~~~~~~~~~~~~~~~~~~~~~~~~Fig. 10~~~~~~~~~~~~~~~~~~~~~~~~~~~~~~~~Fig. 11~~~~~~\\
\includegraphics[height=2in, width=2in]{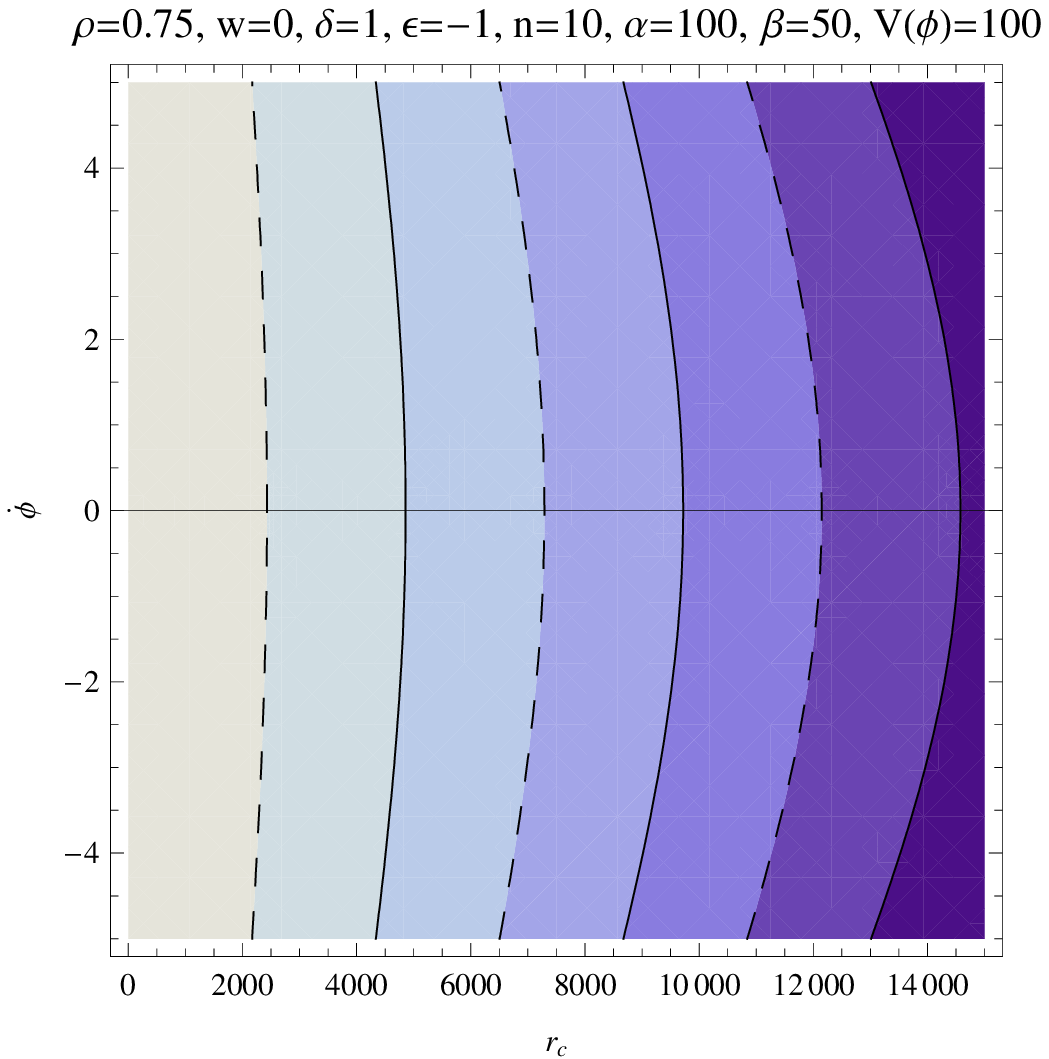}\includegraphics[height=2in, width=2in]{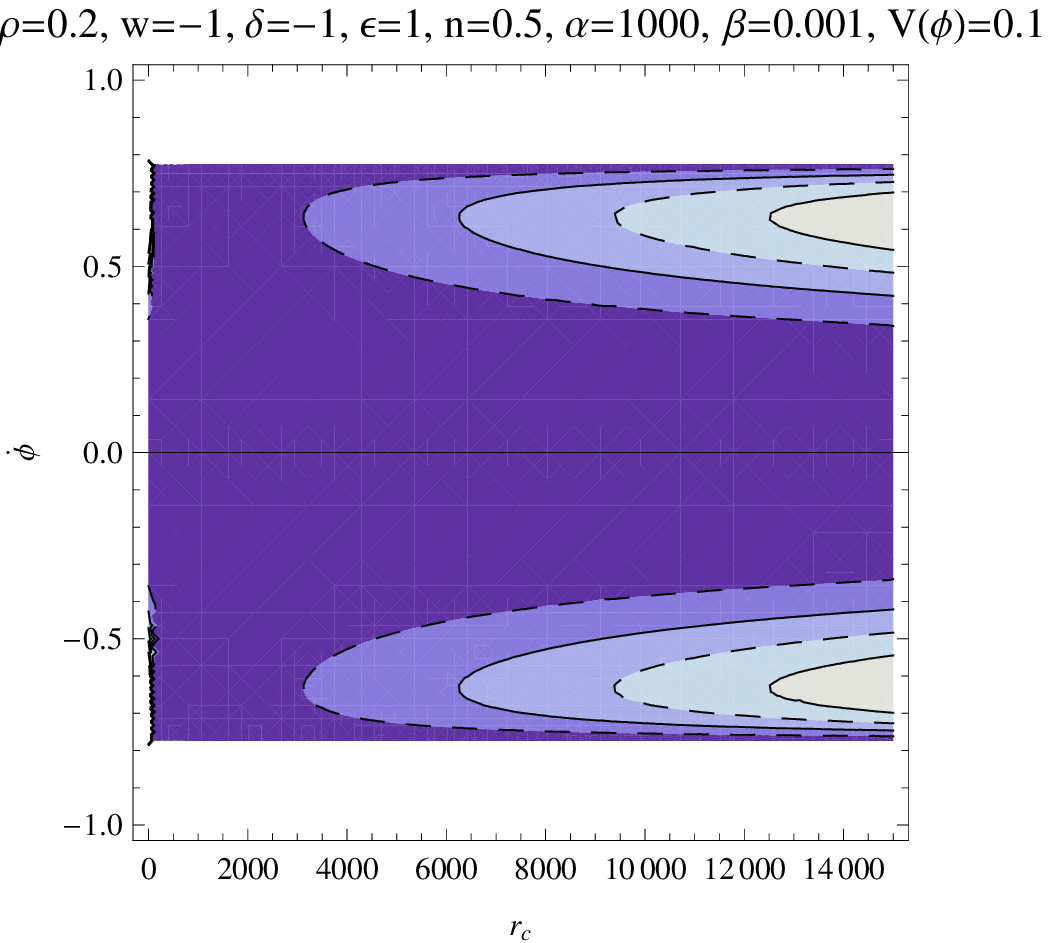}~~\includegraphics[height=2in, width=2in]{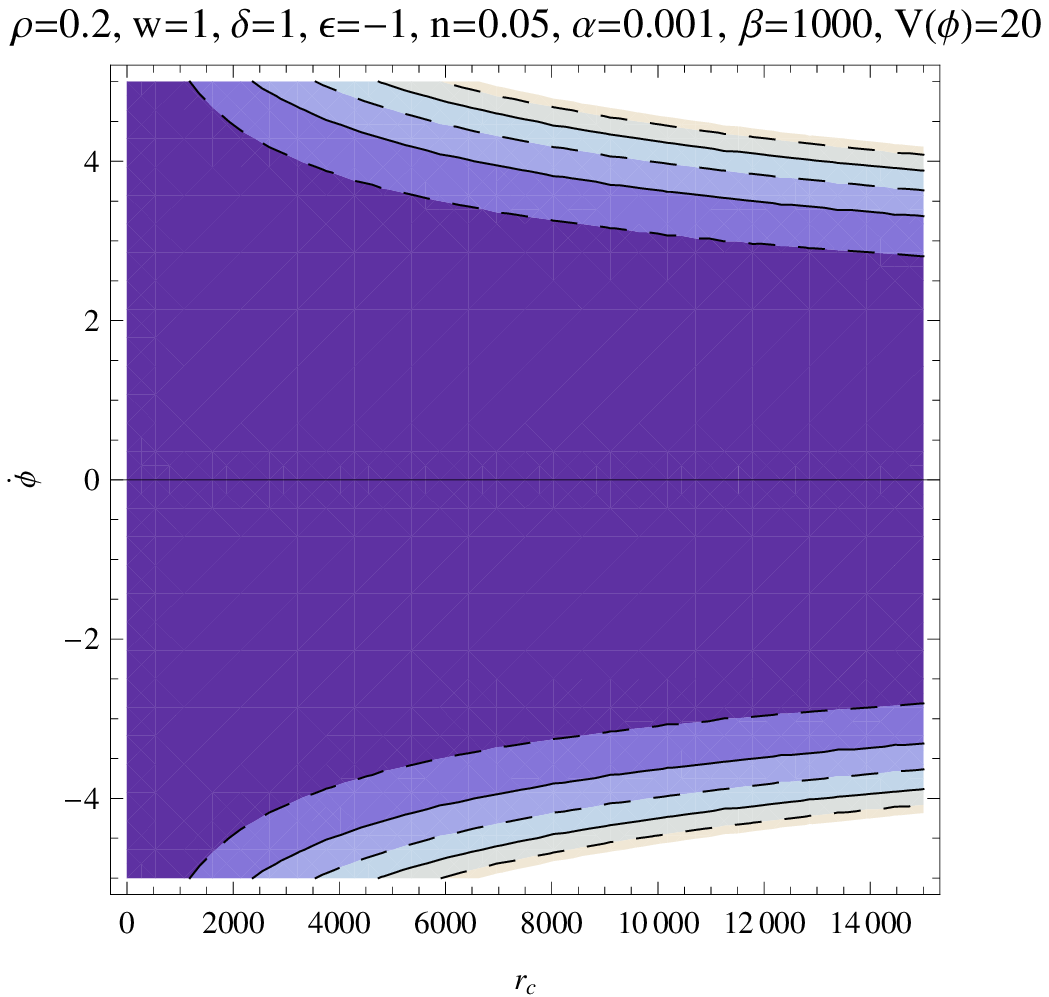}~~\\
\vspace{3mm}
~~~~~~~~~~~~~Fig. 12~~~~~~~~~~~~~~~~~~~~~~~~~~~~~~Fig. 13~~~~~~~~~~~~~~~~~~~~~~~~~~~~~~~Fig. 14~~~~~~\\
\includegraphics[height=2in, width=2in]{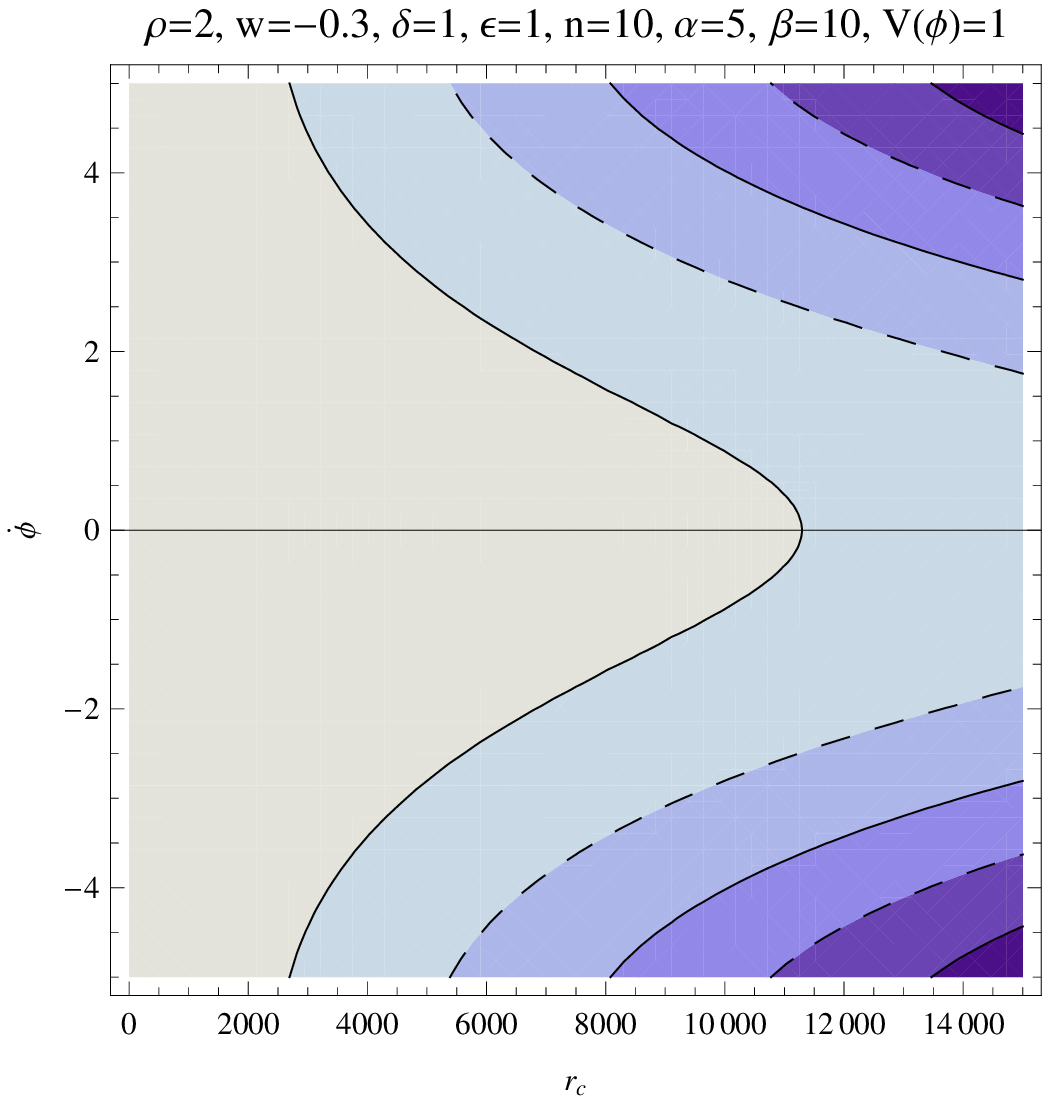}\includegraphics[height=2in, width=2in]{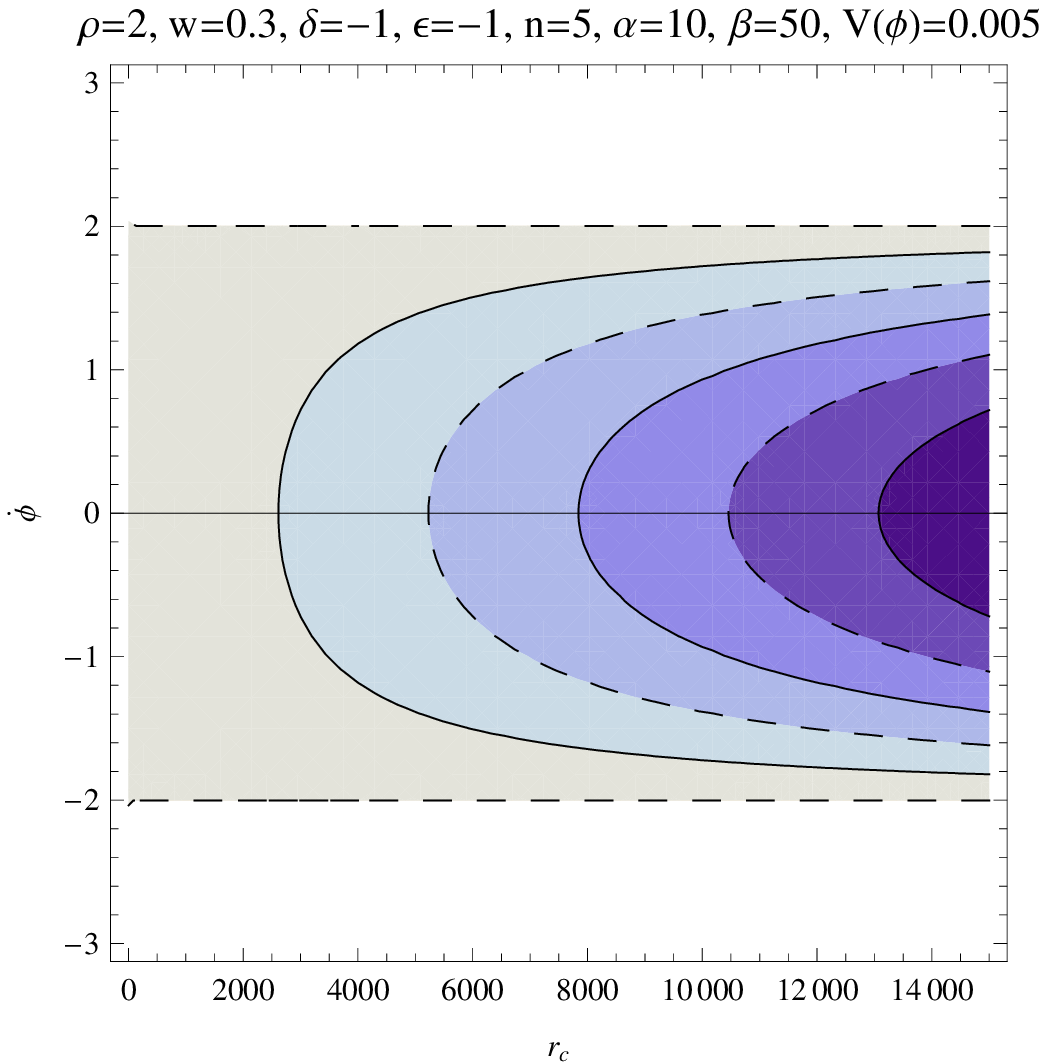}~~\\
\vspace{3mm}
~~~~~~~~~~~~~~Fig. 15~~~~~~~~~~~~~~~~~~~~~~~~~~~~~~Fig. 16~~\\

The above figures represent the plot of $\dot{\phi}$ against
$r_{c}$ (in equation (20)) with the variation of different
parameters. The different values of the parameters chosen have
been provided with the plots.
\end{figure}

\vspace{3mm}

\noindent

From the plots below it can be seen that on most occasions we get
a positive root of $\dot{\phi}$ from equation (20), irrespective
of whatever sets of values we choose for the associated
parameters. Both the branches of DGP brane model has been taken
into account and both normal and phantom fields have been
considered in the plots. In some cases there are some constraints
specially for low values of $n$, very high values of $r_{c}$ and
extremely low values of $V(\phi)$. In these cases positive roots
are difficult to achieve, and hence the realization of the
emergent scenario. Other than these trivial cases we can easily
experience the emergent universe for DGP brane model in normal and
phantom scalar field associated with barotropic fluid as normal
matter.

\begin{center}
\begin{tabular}{|l|}
\hline\hline
~~$\rho_{m}$~~~~~~~~$w$~~~~~~~~~$\delta$~~~~~~~~~~$\epsilon'$~~~~~~~~~$r_{c}$~~~~~~~~~~~~~~~$n$~~~~~~~~~~~~~~$\alpha$~~~~~~~~~~~~$\beta$~~~~~~~~~~~$V(\phi)$~~~~
~~~Positive roots ($\dot{\phi}$) \\ \hline

\\ \hline\hline
\\
~0.5~~~~~~~~-0.5~~~~~~~1~~~~~~~~~~1~~~~~~~~10~~~~~~~~~~~~~~~0.1~~~~~~~~~~~~~2~~~~~~~~~~~~1~~~~~~~~~~~~~2~~~~~~~~~~~~~~~~~~~~~~~$-$~~~~~~~
\\
~0.2~~~~~~~~-0.8~~~~~~~1~~~~~~~~~~1~~~~~~~~100~~~~~~~~~~~~~~0.7~~~~~~~~~~~~~5~~~~~~~~~~~0.01~~~~~~~~~~0.01~~~~~~~~~~~~~~~0.407962~~~~
\\
~0.01~~~~~~~-0.1~~~~~~~1~~~~~~~~~~1~~~~~~~~0.1~~~~~~~~~~~~~~~1~~~~~~~~~~~~~10~~~~~~~~~~~100~~~~~~~~~~~100~~~~~~~~~~~~~~~~1.76366~~~~
\\\hline\hline\\
~5~~~~~~~~~~~0.5~~~~~~~-1~~~~~~~~~-1~~~~~~~10~~~~~~~~~~~~~~~1.2~~~~~~~~~~~~~~2~~~~~~~~~~~10~~~~~~~~~~~~50~~~~~~~~~~~~~~~~~~2.90562~~~~
\\
~0.05~~~~~~~0.4~~~~~~~-1~~~~~~~~~-1~~~~~~~100~~~~~~~~~~~~~~1.5~~~~~~~~~~~~~10~~~~~~~~~0.001~~~~~~~~1000~~~~~~~~~~~~~~~~0.424633~~~
\\
~0.005~~~~~~0.3~~~~~~~-1~~~~~~~~~-1~~~~~~~1000~~~~~~~~~~~~~~2~~~~~~~~~~~~~100~~~~~~~~~0.1~~~~~~~~~~100~~~~~~~~~~~~~~~~~14.1425~~~
\\
~70~~~~~~~~~0.2~~~~~~~-1~~~~~~~~~-1~~~~~~~~10000~~~~~~~~~~~10~~~~~~~~~~~~~0.1~~~~~~~~~100~~~~~~~~~~~0.1~~~~~~~~~~~~~~~~~~~~$-$~~~
\\
~0.75~~~~~~~0.7~~~~~~~-1~~~~~~~~~-1~~~~~~~~0.01~~~~~~~~~~~~100~~~~~~~~~~~~0.01~~~~~~~~50~~~~~~~~~~~~10~~~~~~~~~~~~~~~~~~4.57297~~~~~
\\\hline\hline\\
~0.25~~~~~~~0.9~~~~~~~1~~~~~~~~~-1~~~~~~~~~~1~~~~~~~~~~~~~~~1~~~~~~~~~~~~~~~2~~~~~~~~~~~0.1~~~~~~~~~~~5~~~~~~~~~~~~~~~~~~~~0.61914~~~~~
\\
~0.25~~~~~~~0.1~~~~~~~1~~~~~~~~~-1~~~~~~~~~10~~~~~~~~~~~~~~1.5~~~~~~~~~~~~~10~~~~~~~~~~5~~~~~~~~~~~~~0.1~~~~~~~~~~~~~~~~~~8.31793~~~
\\
~0.5~~~~~~~~~0~~~~~~~~-1~~~~~~~~~~1~~~~~~~~~100~~~~~~~~~~~~~~5~~~~~~~~~~~~~~50~~~~~~~~~~100~~~~~~~~~~50~~~~~~~~~~~~~~~~~~10.0499~~~~~
\\
~0.75~~~~~~~~0~~~~~~~~1~~~~~~~~~-1~~~~~~~~~1000~~~~~~~~~~~~10~~~~~~~~~~~~100~~~~~~~~~~50~~~~~~~~~~~100~~~~~~~~~~~~~~~~~~~21.2419~~~
\\\hline\hline\\
~0.2~~~~~~~~-1~~~~~~~-1~~~~~~~~~~1~~~~~~~~~~0.2~~~~~~~~~~~~~0.5~~~~~~~~~~~~1000~~~~~~~~0.001~~~~~~~~0.1~~~~~~~~~~~~~~~~$4.32136\times10^{-21}$~~~
\\
~0.2~~~~~~~~~1~~~~~~~~1~~~~~~~~~-1~~~~~~~~~~0.2~~~~~~~~~~~~~0.05~~~~~~~~~~~0.001~~~~~~~1000~~~~~~~~2000~~~~~~~~~~~~~~~~~~~~~$-$~~~
\\\hline\hline\\
~2~~~~~~~~~-0.3~~~~~~~1~~~~~~~~~~1~~~~~~~~~~0.5~~~~~~~~~~~~~~10~~~~~~~~~~~~~~5~~~~~~~~~~~~10~~~~~~~~~~~~1~~~~~~~~~~~~~~~~~~6.66419~~~
\\\hline\hline\\
~2~~~~~~~~~~0.3~~~~~~-1~~~~~~~~~-1~~~~~~~~~~10~~~~~~~~~~~~~~~5~~~~~~~~~~~~~~~10~~~~~~~~~~50~~~~~~~~~~~0.005~~~~~~~~~~~~~~2.00624~~~~
\\
\\ \hline\hline
\end{tabular}
\end{center}
\vspace{8mm}

{\bf Table I:} Nature of the roots ($\dot{\phi}$) of the equation (20) for various values of parameters involved.\\
\vspace{8mm}

Above we have prepared a table showing the positive roots of
$\dot{\phi}$ for different sets of values of the parameters. Both
self-accelerating ($\epsilon'=1$) and non-self accelerating
($\epsilon'=-1$) branches of DGP brane model has been considered.
Moreover both Normal scalar field ($\delta=1$) and phantom scalar
($\delta=-1$) have been considered while finding the roots. From
the table it is seen that in general there are no positive roots
of $\dot{\phi}$ for very small value of $n$. For very high value
of $r_{c}$ as well we are deprived of a positive root. Hence in
these cases there is some difficulty in realizing the emergent
scenario. So it can be concluded that $n$ and $r_{c}$ are the two
really sensitive quantities that govern the realization of an
emergent scenario in the DGP brane model. The dependence on
$r_{c}$ is quite obvious, since it is the cross over scale from
$4D$ to $5D$. A very large value of $r_{c}$ definitely alters the
basic nature of the model and may be responsible for any sort of
adverse phenomenon. But the dependence on $n$ is not quite clear
and remains an intriguing question. Perhaps its appearance in the
numerator of the expression of Hubble parameter and its
derivatives, gives it enough independence to play a major role in
determining the fate of the universe. But barring these special
cases we are definitely able to realize the emergent scenario for
DGP brane model on a general basis.

\subsection{TACHYONIC FIELD}

Using the field equation for DGP brane, i.e., equations
(\ref{emergent6.15}) and (\ref{emergent6.16}) with the equations
(\ref{emergent6.12}) and (\ref{emergent6.13}) we get,
\begin{equation}
\dot{\psi}^{2}=\frac{\rho_{m}\left(1+w\right)\left(\sqrt{\frac{4r_{c}^{2}\rho+3}{3}}+\epsilon'\right)-2\dot{H}\sqrt{\frac{4r_{c}^{2}\rho+3}{3}}}{\left(\sqrt{\frac{4r_{c}^{2}\rho+3}{3}}+\epsilon'\right)\rho_{\psi}\epsilon}
\end{equation}
In order to make the R.H.S. of the above equation positive
definite we should have,
\begin{equation}
\dot{H}<\frac{\rho_{m}\left(1+w\right)\left(\alpha+\epsilon'\right)}{2\alpha}
\end{equation}
where $\alpha=\sqrt{\frac{4r_{c}^{2}\rho+3}{3}}$. The above
relation gives the condition for the emergent universe with
Tachyonic field in DGP brane model. Now in order to be consistent
with the recent cosmic acceleration the above relation for the
emergent universe gives the following range for $\dot{H}$,
$\dot{H}\in\left(0,\frac{\rho_{m}\left(1+w\right)\left(\alpha+\epsilon'\right)}{2\alpha}\right)$.

%%%%%%%%%%%%%%%%%%%%%%%%%%%%%%%%%%%%%%%%%%%%%%%%%%%%%%%%%%
\section{EMERGENT SCENARIO IN KALUZA-KLEIN COSMOLOGY}
%%%%%%%%%%%%%%%%%%%%%%%%%%%%%%%%%%%%%%%%%%%%%%%%%%%%%%%%%%

\noindent

\vspace{3mm}

Einstein's field equation for Kaluza-Klein space-time is given by,
\begin{equation}\label{emergent6.18}
H^{2}+\frac{k}{a^{2}}=\frac{4\pi G}{3}\rho
\end{equation}
and
\begin{equation}\label{emergent6.19}
\dot{H}+2H^{2}+\frac{k}{a^{2}}=-\frac{8\pi G}{3}p
\end{equation}

\vspace{3mm}

%%%%%%%%%%%%%%%%%%%%%%%%%%%%%%%%%%%%%%%%%%%%%%%
\subsection{NORMAL OR PHANTOM FIELD:}
%%%%%%%%%%%%%%%%%%%%%%%%%%%%%%%%%%%%%%%%%%%%%%%

\noindent

\vspace{3mm}

Using equations (\ref{emergent6.18}) and (\ref{emergent6.19}) we
have,
\begin{equation}\label{emergent6.20}
\dot{H}-\frac{k}{a^{2}}=-\frac{8\pi G}{3}\left(p+\rho\right)
\end{equation}
Substituting the values of $\rho$ and $p$ from equations
(\ref{emergent 6.2.1}), (\ref{emergent6.6}) and
(\ref{emergent6.7}) in the above equation we get,
\begin{equation}\label{emergent6.21}
\frac{8\pi
G}{3}\left\{\delta\dot{\phi}^{2}+\rho_{m}\left(1+w\right)\right\}+\left(\dot{H}-\frac{k}{a^{2}}\right)=0
\end{equation}
From the above equation, we get,
\begin{equation}\label{emergent6.22}
\dot{\phi}^{2}=\frac{1}{\delta}\left[\frac{3\left(k-a^{2}\dot{H}\right)}{8a^{2}\pi
G}-\rho_{m}\left(1+w\right)\right]
\end{equation}

\vspace{3mm}

%%%%%%%%%%%%%%%%%%%%%%%%%%%%%%%%%%%%%%%%%%%%%%%%%%%%%%%%%%%%%%
\subsubsection{Case-I: For Normal Scalar field}
%%%%%%%%%%%%%%%%%%%%%%%%%%%%%%%%%%%%%%%%%%%%%%%%%%%%%%%%%%%%%%

\noindent

\vspace{3mm}

For this case $\delta=1$. Hence putting this in equation
(\ref{emergent6.22}) we get,
\begin{equation}\label{emergent6.23}
\dot{\phi}^{2}=\frac{3\left(k-a^{2}\dot{H}\right)}{8\pi
Ga^{2}}-\rho_{m}\left(1+w\right)
\end{equation}
From the above equation we see that the L.H.S is non-negative.
Hence in order that the R.H.S. is also non-negative we should have
\begin{equation}\label{emergent6.24}
\rho_{m}<\frac{3\left(k-a^{2}\dot{H}\right)}{8\pi
Ga^{2}\left(1+w\right)}
\end{equation}
Depending on the above value, we discuss the possibilities of all
the three types of universe.

\vspace{3mm}

$$\bf Subcase-I: For ~Closed ~Universe, ~(k>0)$$

\noindent

\vspace{3mm}

In equation (\ref{emergent6.24}), since the L.H.S. is positive,
hence we should have $\dot{H}<\frac{k}{a^{2}}$. Since $k$ is
positive we can realize an emergent scenario in the range
$\dot{H}\in \left(0, \frac{k}{a^{2}}\right)$.

\vspace{3mm}

$$\bf Subcase-II: For ~Open ~Universe, ~(k<0)$$

\noindent

\vspace{3mm}

For this case we see that since $k<0$, we should have $\dot{H}<0$
in order to make the R.H.S of equation (\ref{emergent6.24})
positive. But this result is contrary to cosmic acceleration,
which is unacceptable.

\vspace{3mm}

$$\bf Subcase-III: For ~Flat ~Universe, ~(k=0)$$

\noindent

\vspace{3mm}

In this case $k=0$. Using this in equation (\ref{emergent6.24}),
we get $\dot{H}<0$, just like the previous case. Hence this case
is also un-physical.

\vspace{3mm}

So the only physically acceptable situation is the first case,
which is the closed universe.

\vspace{3mm}

%%%%%%%%%%%%%%%%%%%%%%%%%%%%%%%%%%%%%%%%%%%%%%%%%%%
\subsubsection{Case-II: For Phantom field}
%%%%%%%%%%%%%%%%%%%%%%%%%%%%%%%%%%%%%%%%%%%%%%%%%%%

\noindent

\vspace{3mm}

For this case $\delta=-1$. Hence putting this in equation
(\ref{emergent6.22}) we get,
\begin{equation}\label{emergent6.26}
\dot{\phi}^{2}=\rho_{m}\left(1+w\right)-\frac{3\left(k-a^{2}\dot{H}\right)}{8\pi
Ga^{2}}
\end{equation}
Here in order to make the R.H.S. positive definite, we should have
\begin{equation}\label{emergent6.27}
\rho_{m}>\frac{3\left(k-a^{2}\dot{H}\right)}{8\pi
Ga^{2}\left(1+w\right)}
\end{equation}
We discuss the different possibilities below:

\vspace{3mm}

$$\bf Subcase-I: For ~Closed ~Universe, ~(k>0)$$

\noindent

\vspace{3mm}

In this case $\dot{H}\leq\frac{k}{a^{2}}\Rightarrow\dot{H}>0$ in
the range $\dot{H}\in \left(0, \frac{k}{a^{2}}\right)$, since
$k>0$. Hence we can realize the emergent scenario in this range,
consistent with the cosmic acceleration.

\vspace{3mm}

$$\bf Subcase-II: For ~Open ~Universe, ~(k<0)$$

\noindent

\vspace{3mm}

Just like the Normal scalar field we do not realize the emergent
universe in this case.

\vspace{3mm}

$$\bf Subcase-III: For ~Flat ~Universe, ~(k=0)$$

\noindent

\vspace{3mm}

In this case, like the normal scalar field we get $\dot{H}<0$.
Hence this case is also un-physical and the emergent scenario is
out of question.\\

Hence here also the emergent scenario is realized only for closed
universe.

\vspace{3mm}

\subsection{TACHYONIC FIELD}

\noindent

\vspace{3mm}

Using equations (\ref{emergent6.13}) and (\ref{emergent6.19}), we
get,
\begin{equation}\label{emergent6.28}
\left[\frac{\frac{3}{8\pi G
}\left(\dot{H}+2H^{2}+\frac{k}{a^{2}}\right)-w\rho_{m}}{U(\psi)}\right]^{2}=1-\epsilon\dot{\psi}^{2}
\end{equation}

\noindent

\vspace{3mm}

L.H.S. of the above equation is positive definite, and this is
possible for all values of $k$, i.e., for open, closed and flat
model of the universe. So it is possible to have emergent scenario
for any model of universe(open, closed or flat) with Tachyonic
field as matter content.

\vspace{3mm}

%%%%%%%%%%%%%%%%%%%%%%%%%%%%%%%%%%%%%%%%%%%%%%%%%%%%%%%%%%%%%%%%%%%%%
\section{CONCLUSION}
%%%%%%%%%%%%%%%%%%%%%%%%%%%%%%%%%%%%%%%%%%%%%%%%%%%%%%%%%%%%%%%%%%%%%

\noindent

\vspace{3mm}

In this paper we have investigated the conditions that help to
realize the emergent scenario of the universe. Three different
models of universe were considered for the discussion, namely,
Loop quantum cosmology, DGP brane-world and Kaluza-Klein
cosmology. Normal, Phantom and Tachyonic fields have been
considered separately for each model. In Tachyonic field, both
normal tachyon and phantom tachyon has been considered for our
evaluation process. To complement the scalar fields barotropic
fluid has been chosen as the normal matter content. In case of
Loop quantum cosmology the emergent scenario is realized with a
restriction on the value of the density of normal matter,
$\rho_{m}$ for normal and phantom field, which is really an
unexpected result as the dominating matter component is the scalar
field. For Tachyonic field emergent scenario is realized with
constraints on the value of $\rho_{1}$ for both normal and phantom
tachyon which is quite understandable. It is found that the
realization of emergent scenario in case of DGP brane model is
possible almost unconditionally in most cases for normal and
phantom scalar fields except in a few exceptional cases. Plots
were generated in order to confirm this. A table containing useful
data has also been provided. In case of tachyonic field we get a
constraint on $\dot{H}$. In case of Kaluza-Klein cosmology
emergent scenario is realized only for closed universe in case of
normal and phantom scalar fields. Finally with tachyonic field as
matter content emergent scenario is realized for all the
models(closed, open, flat) of the universe irrespective of any
conditions.


\vspace{3mm}

\begin{thebibliography}{99}
\small

\bibitem{Ellis1} Ellis, G. F. R., Maartens, R. {\it Class. Quant. Grav.} {\bf 21}, 223 (2004).
\bibitem{Ellis2} Ellis, G. F. R., Murugan, J., Tsagas, C. G. {\it  Class. Quant. Grav.} {\bf 21}, 233 (2004).
\bibitem{Perlmutter1} Perlmutter, S. et al. :- {\bf [Supernova Cosmology Project Collaboration]}, {\it ApJ} {\bf 517}, 565(1999) {\it [arXiv:astro-ph/9812133]}.
\bibitem{Spergel1} Spergel, D. N. et al. :- {\bf WMAP Collaboration},{\it Astron. J. Suppl} {\bf 148}, 175(2003) {\it [arXiv :astro-ph/0302209]}.
\bibitem{Gibbons1} Gibbons, G. W. :- {\it Nucl. Phys. B} {\bf 292}, 784 (1987);
\bibitem{Gibbons2} Gibbons, G. W. :- {\it Nucl. Phys. B} {\bf 310}, 636 (1988).
\bibitem{Mukherjee1}  Mukherjee, S., Paul, B. C., Dadhich, N., K., Maharaj, S. D., Beesham, A. : {\it  Class. Quantum. Grav.} {\bf 23}, 6927(2006).
\bibitem{Debnath1} Debnath, U. : {\it  Class. Quantum. Grav.} {\bf 25}, 205019(2008).
\bibitem{Campo1} S. del Campo et al (2007) JCAP 11 030.
\bibitem{Mukherji1} S. Mukerji and S. Chakraborty (2010), Int. J. Theor. Phys. 49 2446.
\bibitem{Paul1} B. C. Paul et al(2010), Mon. Not. R. Astron. Soc. 407, 415419.
\bibitem{Mukerji1}  Mukerji, S., Chakraborty, S. {\it Astrophys. Space Sc.} DOI: 10.1007/s10509-010-0456-1 (2010)
\bibitem{Rovelli1} Rovelli, C. :- {\it liv. Rev. Rel.} {\bf 1}1(1998)
\bibitem{Ashtekar1} Ashtekar, A., Lewandowski, J. :- {\it  Class. Quantum. Grav.} {\bf 21}R53(2004)
\bibitem{Ashtekar2} Ashtekar, A. :- {\it  AIP Conf. Proc.} {\bf 861}3(2006)
\bibitem{Bojowald1} Bojowald, M. :- {\it liv. Rev. Rel.} {\bf 8}11(2005)
\bibitem{Ashtekar3} Ashtekar, A. et al :- {\it  Adv. Theor. Math. Phys.} {\bf 7}233(2003)
\bibitem{Kaluza1} Kaluza, T. :- {\it Preus. Acad. Wiss} {\bf F1} 9669(1921)
\bibitem{Klein1} Klein, O. :- {\it A. Phys} {\bf 37} 895 (1926)
\bibitem{Polchinski1} Polchinski, J. :- String Theory, Vols. I and II (Cambridge Univ. Press, 1998).
\bibitem{Duff1} Duff, M. J., Nilsson, B. E. W., Pope, C. N. :- {\it Phys. Rep.} {\bf 130} 1 (1986).
\bibitem{Green1} Green, M. B., Schwarz, J. H., Witten, E. :- Superstring Theory (Cambridge Univ. Press, 1987).
\bibitem{Li1} Li, L. X., Gott, I., Richard, J. :- {\it Phys. Rev. D} {\bf 58} 103513 (1998).
\bibitem{Overduin1} Overduin, J. M., Wesson, P., S. :- {\it Phys. Rep.} {\bf 283} 303 (1997).
\bibitem{Adhav1} Adhav, K. S., Nimkar, A. S., Dawande, M. V., :- {\it Int. J. Theor. Phys.} {\bf 47} 2002 (2008).
\bibitem{Qiang1} Qiang, L. E., Ma, Y. G., Han, M. X., Yu,  D. :- {\it Phys. Rev. D} {\bf 71} 061501 (2005).
\bibitem{Chen1} Chen, S., Jing, J. :- {\it J. Cosmol. Astropart. Phys.} {\bf 09} 001(2009).
\bibitem{Ponce de Leon1} Ponce de Leon, J. :- {\it Gen. Relativ. Gravit.} {\bf 20} 539 (1988).
\bibitem{Chi1} Chi, L. K. :- {\it Gen. Relativ. Gravit.} {\bf 22} 1347 (1990).
\bibitem{Fukui1} Fukui, T. :- {\it Gen. Relativ. Gravit.} {\bf 25} 931 (1993).
\bibitem{Liu1} Liu, H., Wesson, P. S. :- {\it Int. J. Mod. Phys. D} {\bf 3} 627 (1994).
\bibitem{Coley1} Coley, A. A. :- {\it Astrophys. J.} {\bf 427} 585 (1994).

\bibitem{Wu1} Wu, P., Zhang, S.N. : {\it J. Cosmol. Astropart. Phys.} {\bf 06}, 007 (2008)
\bibitem{Chen2} Chen, S., Wang, B., Jing, J. : {\it Phys. Rev. D} {\bf 78}, 123503 (2008)
\bibitem{Fu1} Fu, X., Yu, H., Wu, P. : {\it Phys. Rev. D} {\bf 78}, 063001 (2008)


\bibitem{Chang1} Chang, B., Liu, H., Xu, L., Zhang, C. : {\it Chin. Phys. Lett.} {\bf 24} 2153(2007) [{\it arXiv}:{\bf 0704.3768}(astro-ph)]
\bibitem{Hao1} Hao, J. G., Li, X. -z. : {\it Phys. Rev. D} {\bf 68}, 043510 (2003)
\bibitem{Hao2} Hao, J. G., Li, X. -z. : {\it Phys. Rev. D} {\bf 68}, 083514 (2003)
\bibitem{Nojiri1} Nojiri, S., Odintsov, S. D. : {\it Phys. Lett. B} {\bf 571},1 (2003)
\bibitem{Gumjudpai1} Gumjudpai, B., Naskar, T., Sami, M., Tsujikawa, S., : {\it JCAP} {\bf 0506} 007 (2005)
\bibitem{Dvali1}Dvali, G. R. , Gabadadze, G., Porrati, M.:- {\it  Phys.Lett. B} {\bf 485} 208(2000)[{\it arXiv}:{\bf 000506}(hep-th)].
\bibitem{Deffayet1} Deffayet, D.:- {\it  Phys.Lett. B} {\bf 502} 199(2001).
\bibitem{Deffayet2} Deffayet, D., Dvali, G.R., Gabadadze, G.:-{\it  Phys.Rev.D} {\bf 65} 044023 (2002)[{\it arXiv}:{\bf 0105068}(astro-ph)].
\end{thebibliography}
\end{document}